\documentclass[epj]{svjour}
\usepackage{graphicx}
\begin{document}
\title{Verification of atomic data for He- and Li-like ions employing $K$-spectra from the tokamak plasma}
\author{F.~Goryaev\thanks{{e-mail:} goryaev\_farid@mail.ru}
\and A.~Urnov
%
}                     
\offprints{}          
\institute{P.N.~Lebedev Physical Institute of the Russian Academy of Sciences, 119991 Moscow, Russia}
\date{Received: date / Revised version: date}
%
\abstract{
X-ray emission {\it K}-spectra of highly charged He- and Li-like argon ions recorded with high spectral, spatial, and temporal resolution by means of a Bragg spectrometer/polarimeter installed at the TEXTOR tokamak were employed to develop a self-consistent approach (SCA) for deriving information on plasma parameters and the verification (i.e., the estimation of the accuracy) of both atomic data, needed for spectra interpretation, and methods of their calculation. This approach is based on solving the spectral inverse problem for these spectra in the framework of the semi-empirical ``spectroscopic model'' (SM) by means of two complimentary inversion methods: fitting procedure (FP) and that based on Bayes's theorem and called Bayesian iterative method (BIM). The SCA was justified by comparing and analyzing measured and synthetic spectra on the basis of the calculated and corrected atomic data. The three different sets of atomic data (spectral and collisional characteristics) for the Ar$^{16+}$ and Ar$^{15+}$ ions were analyzed and verified.
The developed approach for interpreting the experimental results from the TEXTOR tokamak allowed us to verify the methods for calculating the atomic data with an accuracy of $\sim$~5--10\%.
The spectra calculated with corrected atomic data are in agreement with the spectra measured in the wide range of plasma conditions within the experimental accuracy of 10\%.
Furthermore, corrected atomic data made it possible to perform an accurate diagnostics of plasma parameters: plasma temperatures and relative ion abundances in the tokamak plasma.
This procedure provided also a method for determining the temperature of the plasma core with high accuracy to within 5\%, that is in a good agreement with the diagnostics technique based on the electron cyclotron emission (ECE) data.
The values for relative ion abundances obtained by the application of the spectroscopic and impurity transport model are in agreement within the experimental errors. The presented results show that the X-ray spectroscopy of tokamak plasma is an effective tool for both high accuracy verification of atomic data and precision plasma diagnostics.
} 

\titlerunning{Verification of atomic data for He- and Li-like ions employing $K$-spectra}
\authorrunning{F.~Goryaev}
\maketitle
\section{Introduction}
\label{intro}
High resolution X-ray spectroscopy of highly charged ions of medium nuclear charge, $Z$, is routinely used to determine such important parameters of the tokamak plasma as the central ion temperature $T_{i}$ and toroidal velocity $v_{i}$. Since the first measurements of $T_{i}$ at the Princeton Large Torus (PLT) \cite{Bit79a,Bit79b}, and the Tokamak Fontenay-aux-Roses (TFR) \cite{TFR81a}, crystal spectroscopy is widely applied in present tokamak experiments as an effective diagnostics tool complementary to the Charge Exchange Recombination Spectroscopy (CXRS) \cite{Isl84,Fon83} and is considered as one of the main diagnostics on future large tokamaks \cite{Bit93,Ber99}. In addition to the Doppler shifts of emission lines, caused by poloidal plasma rotation, the X-ray spectra also contain important information on physical processes, plasma dynamics, and plasma parameters (such as electron temperature $T_{e}$, relative
ion abundances $n_{z}$, characteristics of plasma transport and others). The main principles of spectroscopic diagnostics methods for hot low density (coronal) plasma based on relative intensities of He-like ion lines and their dielectronic satellites were developed more than three decades ago for astrophysical applications (see, e.g., \cite{Gab69,Pre79}). Later, these methods have been considerably elaborated and adopted to diagnostics of fusion devices such as JET \cite{Bom88}, TFR \cite{TFR85}, Tore Supra \cite{Pla99}, TEXTOR \cite{Wei01}, FTU \cite{Lei02}, and NSTX \cite{Bit03}. The spectra of highly charged He-like ions abundant in space plasma are extensively used and have to become an indispensable tool in future experiments for a study of physical properties of hot plasma structures in astrophysical objects (such as supernova remnants, accumulation of galaxies, stellar and solar coronae). Similar techniques of the X-ray spectroscopy have also found a diagnostics use in various fundamental and applied studies employing hot plasmas for a creation of the sources of X-ray beams, lithography, material science, and other domains of the modern science and technology.

For the last decades, {\it K}-spectra of He-like ions and their dielectronic satellites, associated with the transitions $nl\to 1s$ of the optical electron, have been effectively used for measuring parameters of hot plasmas. These spectra emitted from low density plasma have been studied previously in a number of works (see, e.g., \cite{Bom88}-\cite{Smi00} for tokamak and \cite{Kee86,Phi93} for solar spectra). The importance of such investigations for plasma diagnostics as well as for the
physics of highly ionized atoms was also pointed out many times (see, e.g., \cite{Kae88}-\cite{Smi00}, \cite{Whi02}). These and other works provided the understanding of the relative roles for the main physical processes responsible for the formation of the spectra. However, in spite of a good agreement achieved between \emph{ab initio} calculated synthetic spectra and experimental ones (see, e.g., \cite{TFR85}), a series of substantial discrepancies in predicted and measured line intensities as well as in wavelengths were reported in many of the aforementioned references (for instance, in \cite{Ber99}, \cite{Ros99}, \cite{Smi96}, \cite{Kee86}). Besides, such an approach does not allow to distinguish the reasons for these discrepancies, whether these are caused by the errors in atomic data or those in plasma characteristics used for the modeling (diffusion coefficients, convective velocities, electron and ion temperatures, ionic abundances). Difficulties in the quantitative estimation of errors in atomic data also arise when they are collected by compilation and/or extrapolation of the results of calculations by different methods not consistent with each others and thus being the source of additional uncertainties in the analysis of their accuracy.

At the same time, the efficiency of spectroscopic methods, and even the very possibility of their application (e.g., in polarization measurements \cite{Dub96}), is critically dependent on the precision of atomic data used for spectra modeling. Hence the experimental verification of such data providing a knowledge on their accuracy, besides the fundamental importance for atomic physics, is badly needed for unambiguous description of the mechanisms of spectra formation and precision determination of appropriate plasma parameters. It is also necessary to develop advanced diagnostic techniques, for example, for future investigations of fast and non-Maxwellian phenomena in hot plasmas.

On the one hand, direct measurements of spectroscopic and collisional characteristics of highly charged ions by crossed-beam experiments, are now practically absent. Thus the only source of information on their accuracy are the same {\it K}-spectra recorded from low density plasma or beam-plasma experiments. Due to narrow spectral lines, the EBIT sources are traditionally used to measure wavelengths, life-times of metastable states and electron-ion cross-sections (see, e.g., \cite{Lev88,Bei96}). However, they are not always
suitable for verification of collisional data; an accuracy of 20--30\% or less is far not sufficient for many diagnostic purposes.

On the other hand, being characterized by a high photon flux intensity, the {\it K}-spectra of tokamak plasma recorded consistently with other diagnostics techniques, provide a unique possibility to test the atomic data calculated by various methods. As was shown in \cite{Urn07}, the correction and analysis of atomic data accuracy simultaneously with plasma characteristics (temperatures of plasma core) can be done by means of a self-consistent approach based on a solution of the inverse problem for these spectra in the framework of the semi-empirical ``spectroscopic model'' (SM).

The present work is one of the steps in a wide program launched earlier in \cite{Ber99,Wei01,Urn07,Ros99,Gor03,Mar06a,Mar06b} and aimed at precision studies of atomic and plasma processes by investigating highly charged argon ions spectra by means of the X-ray spectroscopy of the Torus Experiment for Technology Oriented Research (TEXTOR) tokamak plasma. It is devoted to an application of the self-consistent approach to the problem of verification by means of the Ar$^{16+}$ and Ar$^{15+}$ spectra in the range 3.94--4.02~\AA\ obtained with high spectral, spatial, and temporal resolution.

In contrast to previous works the present paper deals with the quantitative verification of the unified methods used for the calculations of all atomic data rather than the individual characteristics needed for {\it K}-spectra description. For this
purpose the improved spectroscopic and collisional data for He- and Li-like argon ions have been calculated by atomic physics codes developed at the Lebedev Physical Institute and the Paris Observatory, corresponding to the atomic physics approaches based on the perturbation theory and multi-configuration expansions, respectively. The analysis using the relative flux intensities gave insight into atomic processes contributing to the observed spectra and helped to understand the mechanisms of the plasma physics involved. A correction of atomic characteristics resulted in the modified wavelengths and effective rate coefficients recommended for spectroscopic diagnostics of low density hot
plasmas containing argon. Our main purpose is to show that a precision of the measured values for a key atomic data, actually stipulating the spectra of hot plasmas, noticeably exceed the accuracy of previous experiments and theoretical estimations. The synthetic spectra based on these data provide an accurate description of the measured {\it K}-spectra for the whole identified spectral region at substantially different  electron temperatures and densities of the tokamak plasmas. This result allows the measurements of corresponding plasma parameters with high accuracy necessary for an application of advanced diagnostic techniques mentioned above.

The outline of this paper is as follows. The formulation of the verification problem employing the {\it K}-spectra is given in Sect. 2. X-ray spectroscopy in the experiment at the TEXTOR tokamak is presented in Sect. 3. In Section 4 the spectroscopic model is described and two complimentary inversion procedures are considered. The analysis of the results of the modeling and correction of calculated atomic data, wavelengths and collisional characteristics, is presented in Sect. 5. The summary of results and conclusions are given in Sect. 6.

\section{Formulation of the problem}
\label{sec:2}
The line spectrum of Ar$^{16+}$ (He-like) and Ar$^{15+}$ (Li-like) ions under study covers the wavelength range 3.94--4.02~\AA . The most prominent features of the spectra result mainly from the lines caused by transitions in Ar$^{16+}$ ion: resonance $1s 2p(^{1}P_{1})\to 1s^{2}(^{1}S_{0})$, magnetic quadrupole $1s 2p(^{3}P_{2}) \to 1s^{2}(^{1}S_{0})$, intercombination $1s 2p(^{3}P_{1})\, \to \, 1s^{2}(^{1}S_{0})$, and forbidden $1s 2s(^{3}S_{1})\to 1s^{2}(^{1}S_{0})$, designated as \textbf{w}, \textbf{x}, \textbf{y}, and \textbf{z} lines, respectively (these notations are used following \cite{Gab72}). These lines are produced primarily due to direct electron impact excitation including cascades from higher levels as well as due to contributions from radiative and dielectronic recombination of Ar$^{17+}$ (H-like) ions, charge exchange of Ar$^{17+}$ ions with neutral hydrogen atoms, inner-shell ionization of Ar$^{15+}$ ions (contributing to the \textbf{z} line), and resonance scattering via doubly excited autoionizing states $1snln'l'$ ($n,n'>2$).

\begin{figure}
\resizebox{0.5\textwidth}{!}{%
  \includegraphics[angle=270]{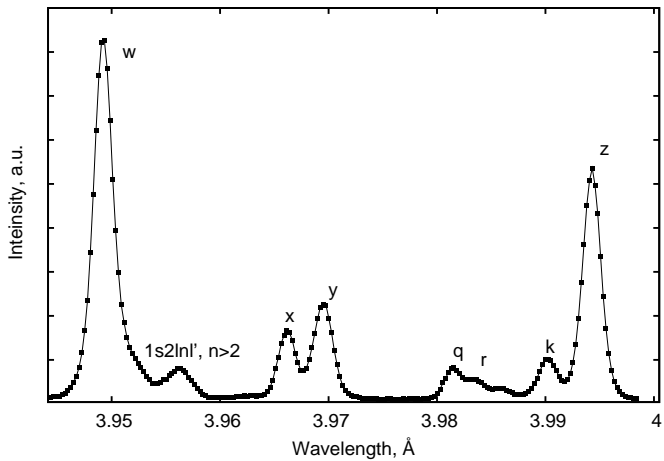}
}
\caption{Example of the measured spectrum of Ar$^{16+}$ and Ar$^{15+}$ ions. The solid line
presents the theoretical spectrum, and the small squares correspond to the measured one (by
courtesy of O. Marchuk).}
\label{fig:1}       
\end{figure}

Additionally, there are numerous spectral lines, called ``dielectronic satellites'' (DS), in the vicinity of He-like ion lines due to transitions $1s 2p nl\to 1s^{2} nl$ in Li-like ions and, to a lesser extent, $1s2s2pnl\to 1s^{2} 2snl$ in Be-like ions. The most prominent DS with $n=2$ resolved in the spectrum are the \textbf{q} and \textbf{r} satellites emitted by the Ar$^{15+}$ ion and excited predominantly by collisions with electrons, and the \textbf{k} and \textbf{a} ones excited by means of the dielectronic recombination mechanism from the Ar$^{16+}$ ion. The most intensive \textbf{j} satellite is blended with the \textbf{z} line. There are also two groups of unresolved DS in the long-wavelength wing of the \textbf{w} line corresponding to dielectronically excited $n=3$ and $n=3,4$ satellites and denoted here as $\mathbf{N}_{3}$ and $\mathbf{N}_{4}$ peaks, respectively. The remaining less-intense satellites densely fill the spectral range, forming series converging to He- and Li-like lines. The typical measured spectra from the TEXTOR tokamak is demonstrated in Fig.~\ref{fig:1}.

\subsection{Self-consistent approach}
\label{subsec:21}
The argon tokamak plasma spectra under consideration consist of a set of well-resolved intensity peaks \{$\mathbf{L}$\}. Each peak $\mathbf{L}$ is in turn associated with the corresponding group of spectral lines \{\textit{l}\} giving a dominant contribution to the total intensity flux of the peak. For example, the peak $\mathbf{Z}$ is mostly formed by the $\mathbf{z}$\ and $\mathbf{j}$ lines, and the contribution of the $\mathbf{z}$ line is dominant at high temperatures. Actually due to a large Doppler broadening at high temperatures each peak has a complex spectral shape presenting a group of blended lines emitted by the same ion (for example, DS) or belonging to different ion species. For the purpose of quantitative analysis of spectral fluxes, the whole spectral range has been divided into the intervals $\mathrm{\Delta} \lambda_{\mathbf{L}}= [\mathbf{L}]$, corresponding to peaks $\mathbf{L}$, and the three spectral regions at the edges and in the middle of the spectrum were chosen for background determination. The following set of peaks was identified: $\{\mathbf{L}\}=\{\mathbf{W,N_{4},N_{3},X,Y,Q,R,A,K,Z}\}$.

The spectral intensity $I(\lambda )$ in the theoretical (synthetic) spectrum as well as the emission fluxes $F_{[\mathbf{L}]} = \int_{[\mathbf{L}]} I(\lambda ) d\lambda$ of the peaks \{$\mathbf{L}$\} in the corresponding spectral intervals $[\mathbf{L}]$ can be considered as functions (or functionals) of the two sets of physical characteristics: (i) atomic data $\mathrm{AD}=\{\mathrm{A}_0,\mathbf{A}(T_e)\}$, and (ii) plasma parameters $\mathrm{PP}=\{\mathrm{P}_0,\mathbf{P}(r)\}$:

\begin{equation}
I(\lambda )=I(\lambda ;\mathrm{AD},\mathrm{PP})\, , \,\,\, F_{[\mathbf{L}]}=F([L];\mathrm{AD},\mathrm{PP})\, .  \label{functional}
\end{equation}
Here $\mathrm{A}_0=\{\lambda_l ,A_l^{(r)},k_{l},...\}$ stands for the set of atomic constants (wavelengths, radiative transition probabilities, branching ratios, etc.), and $\mathbf{A}(T_{e})=\{C_{l}^{z}(T_{e})\}$ includes collisional characteristics (effective rates for excitation of lines $\{ l \}$ from the ions with charge $z$) for elementary processes in the plasma which are functions of the electron temperature $T_{e}$.
The plasma parameters PP are characterized by both their central values in the tokamak plasma core, $\mathrm{P}_0 =\{P_{\nu}^{(0)}\}$, and the normalized radial profiles $\mathbf{P}(r) =\{\mathbf{P}_\nu(r)\}$, where $\mathbf{P}(0)\equiv 1$. The label
$\nu$ specifies the following plasma parameters: electron, $T_{e}$, and ion, $T_{i}$, temperatures, electron density $N_{e}$, argon ion densities $N_{z}$ for the charges $z$, and neutral atom density of the working gas (hydrogen, deuterium or helium) $N_{g}$. Using these notations, the expression $P_{\nu}(r)=P_{\nu}^{(0)}\cdot\mathbf{P}_{\nu }(r)$ denotes the plasma parameter $\nu$ as a function of radius $r$. For instance, the electron density radial profile is given by $P_{Ne}(r)=N_{e}(r)=N_{e}^{(0)}\cdot\mathbf{P}_{Ne}(r)$.

The intensity spectrum $I(\lambda)$ (in arbitrary units) from the emitting plasma column in the equatorial plane of the tokamak along the minor radius $a$ (see Section \ref{sec:3} below) can be written as
\begin{equation}
I(\lambda )=C\int_{0}^{1} j(P_{\nu }(\rho );\lambda )\, [N_{e}(\rho
)]^{2}\, g(\rho )\, d\rho \, ,  \label{spectral intensity}
\end{equation}
where $\rho=r/a$ is a dimensionless parameter associated with the magnetic surfaces,
$g(\rho){\simeq} 1$ is a factor including the correction for both the
geometry of the emitting plasma volume and asymmetry of plasma profiles
relative to the center of the plasma core.

$j(P_{\nu }(\rho
);\lambda )$ is the local spectral emissivity function (per atom
and electron) which may be expressed through the normalized
population density $n_{z}^{(k)}=N_{z}^{(k)}/N(\mathrm{Ar})$, and the
probability of the radiative transition $A_{l}^{(r)}$ for the line
$l=l(k,k^{\prime })$ due to the transition $k\rightarrow
k^{\prime}$, where $N$(Ar) is the argon density and $N_{z}^{(k)}$ is
the density of the excited state $k$:
\begin{eqnarray}
j(P_{\nu }(\rho );\lambda ) &=& \sum_{l}j_{l}(T_e,N_e,N_g;\rho
)\pounds_{l}(T_{i};\lambda -\lambda _{l})\, , \label{emiss} \\
j_{l} &=& n_{z}^{(k)}A_{l}^{(r)}\, , \nonumber
\end{eqnarray}
where the function $\pounds_{l}(T_{i};\lambda -\lambda_l)$ is the counter describing
the line shape due to natural and Doppler broadening; the
summation in Eq. (\ref{emiss}) is  over all lines $l$ having a
contribution to the flux at the wavelength $\lambda $. The
parameter $C$ in Eq. (2) is a conversion coefficient which can be
defined by the condition of  equality of the measured flux
$F_{[\sigma ]}^{(\exp )}$ and predicted one $F_{[\sigma ]}$ in the
spectral range $\mathrm{\Delta} \lambda =[\sigma ]$:
\begin{equation}
F_{[\sigma ]}=\int_{[\sigma ]}I(\lambda )d\lambda \, .  \label{flux-definition}
\end{equation}
The emission flux for the peak $\mathbf{L}$,
$F_{[\mathbf{L}]}$, is expressed then through the emissivity
function $J_{[\mathbf{L}]}(T_e,N_e,N_g;\rho )$ by the equation
\begin{eqnarray}
F_{[\mathbf{L}]}=C\int_{0}^{1}J_{[\mathbf{L}]}(P_{\nu }(\rho
))[N_{e}(\rho )]^{2}\,g(\rho )\,d\rho \, , \label{peak-flux} \\
J_{[\mathbf{L}]}(P_{\nu }(\rho ))=\int_{[\mathbf{L}]}j(P_{\nu
}(\rho );\lambda )d\lambda \, . \nonumber
\end{eqnarray}
A synthetic emission flux in Eq. (\ref{peak-flux}) depends on a particular model based on equations of atomic and plasma kinetics. For the adopted model, as it follows from the analysis given below, one may define the two sets of its basic or ``key'' parameters \textbf{D}=\{\textbf{D}$_{i}$\} and $\mathbf{\Phi }=\{\mathbf{\Phi }_{i}\}$ from the corresponding sets AD and PP respectively (for example, ratios of line excitation rates or central temperature $T_{e}^{(0)}$ in the plasma core), which identically (with a given accuracy) simulate the fluxes of the synthetic spectrum
$F_{[\mathbf{L}]}^{(\mathrm{syn})}=F_{[\mathbf{L}]}(\mathbf{D},\mathbf{\Phi })$ for experimental conditions under consideration. In the frame of semi-empirical models, key plasma parameters $\mathbf{\Phi }$ have to be independently predetermined (calculated or measured) for a direct ``\emph{ab initio}'' calculations of the synthetic spectra, while for diagnostic purposes they can be derived from the measured spectra by solving the inverse spectroscopic problem with a set of equalities:
\begin{equation}
F_{[\mathbf{L}]}^{(\exp
)}=F_{[\mathbf{L}]}(\mathbf{D},\mathbf{\Phi }) \, . \label{exp=th}
\end{equation}
These conditions bind possible values of model parameters by experimental constraints and formally generate an implicit functional relations between $\mathbf{D}$ and $\mathbf{\Phi }$ sets. We have to stress here that Eqs. (\ref{exp=th}) have to be valid for \emph{all spectra} measured at \emph{various plasma conditions}, in particular electron temperature of the plasma core. Due to a \emph{strong dependence} of key atomic data (effective excitation rates for spectral lines) on the temperature the number of \emph{substantially different equations} for each $\mathbf{[L]}$ is much lager as compared to the number of the key parameters of the model. It imposes a strong conditions on the relations between $\mathbf{D}$ and $\mathbf{\Phi }$ and leads to a restriction of the class for possible solutions of the inverse problem. Written explicitly in the form
\begin{equation}
\mathbf{\Phi }=\mathbf{\Phi }(F_{[\mathbf{L}]}^{(\exp
)},\mathbf{D}) \,\,\, \mathrm{or} \,\,\, \mathbf{D}=\mathbf{D}(F_{[\mathbf{L}]}^{(\exp ) },\mathbf{\Phi })\, ,
\label{parameters}
\end{equation}
these equalities have to be complemented by additional physical constraints. Both sets of
parameters, evidently, should not depend on $\mathbf{[L]}$; the atomic characteristics
$\mathbf{D}$, being atomic fundamentals, should not depend on variation of plasma conditions.
Besides, for example, relative ionic abundances $n_{z}=N_{z}/N(\mathrm{Ar})$, derived from the measured spectra, have to obey the normalization condition
\begin{equation}
n(\rho )=\sum_{z}n_{z}(\rho )=1  \label{continuity eq}
\end{equation}
which (as it will be shown below) may not be satisfied for the values $n_{z}(\rho )$ obtained from experimental spectra by the inversion procedure, if the atomic data are not correct. As a result the set of Eqs. (\ref{parameters}) along with the physical requirements (\ref{continuity eq}) provides the conditions of self-consistency for atomic and plasma parameters which can be used for the solution of the verification problem.

By definition the term ``verify'' means prove to be true or check for accuracy. Evidently the verification of some (calculated) physical quantity can be done if its ``true'' values are known. Thus an accurate direct measurement of this quantity provides a sufficient condition of its verification. However such procedure is not a necessary one and indirect measurements can be also used to check for accuracy. Both characteristics, $\mathbf{D}$ and $\mathbf{\Phi }$, being the variable parameters of the model, may be optimized by means of (\ref{continuity eq}), which are shown in \cite{Urn07} (see also below the subsection ``Bayesian inversion'') to be the necessary and sufficient conditions for these parameters to be self-consistent and thus to be correct correct.

The general concept of the self-consistent approach (SCA) to the verification problem used in the present paper includes several aspects or levels of consistency. The main idea of the approach is connected with the aforementioned condition of ``intrinsic'' consistency of key parameters $\mathbf{D}$ and $\mathbf{\Phi }$ in the framework of the SM (see below Section 4.3). The accuracy of the verification procedure depends on: (i) the accuracy of the experimental data, (ii) the number of spectral features in the selected spectral range in comparison with the number of variable parameters and their sensitivity to the latter, (iii) the number of spectra measured at significantly different conditions. In our particular case of the K$_{\alpha }$-spectrum of Ar ions ten prominent peaks consisting of numerous spectral lines are well resolved in seven arbitrarily selected experimental spectra, measured for a wide range of the central electron temperatures $T_{e}=0.8-2.5$~keV and densities $N_{e}=10^{13}-10^{14}$~cm$^{-3}$.

The SCA also implies a self-consistency of the atomic data within a unified method of their calculations. Therefore the verification of atomic data means at the same time the verification of the corresponding approach, avoiding the compilation (or the extrapolation) of data obtained by different methods. In the present work two main methods (with some modifications provided by different code packages) were used, respectively: the expansion of perturbation theory over inverse nuclear charge {\it Z} ({\it Z} expansion method), and that based on the multi-configuration expansion with scaled model potentials being optimized  through the scaling parameters by a minimization energy procedure. In both methods relativistic corrections are accounted for by a Breit Hamiltonian. The data based on the R-matrix method for calculations of electron-ion impact cross-sections, traditionally
considered as the most accurate one, are also included in the analysis.

Another aspect of the approach concerns the consistency of plasma parameters obtained by the spectroscopic method with those measured by independent diagnostic techniques. It is shown here that using only the normalized profiles $\mathbf{P}_{Te}(r)$, $\mathbf{P}_{Ne}(r)$, obtained from the electron cyclotron emission (ECE) signal and interferometer data (HCN) the central values of electron temperature derived from the spectra are in a good agreement with the absolute ECE measurements. The radial profiles for relative ionic abundances $n_{z}$ derived within the framework of the SM have also to coincide with those derived from the spectra by their \emph{ab initio} simulation in the framework of tokamak plasma models. Such simulation by the impurity transport model (ITM) was made in \cite{Mar06b}. The agreement of the results obtained by two different methods (spectroscopic and plasma models), justifying the accuracy of both, is demonstrated below.

In fact, the SCA being applied to the verification of atomic data can be considered as a ``stimulated selection'' similar to a ``natural selection'' implicitly used in the spectroscopy of astrophysical and laboratory plasmas due to long-term experiences. When successively fitting a same spectrum for various plasma conditions occurring in applications, this allows for selecting ``survived'' atomic data as the recommended ones. The proposed SCA procedure makes it possible to accelerate the process of selection on the basis of the SCA due to the possibility to use a wide range of operating conditions (for example, the central electron temperature $T_{e}^{(0)}$) and independent diagnostic techniques available at the TEXTOR tokamak. Note that the algorithm used in the present work though different has some similarities to the genetic algorithm applied to the analysis of X-ray spectra in \cite{Gol02}.

\subsection{Atomic data calculations}\label{subsec:22}
The atomic data needed for simulating the synthetic spectra were calculated by means of two sets of numerical codes referred below to as LPI (Lebedev Physical Institute) and PO (Paris Observatory). The LPI set consists of the ATOM and MZ codes developed at Lebedev Physical Institute (see, e.g., \cite{She93}). The PO set includes the codes developed at the University College of London (UCL) and partly modified at the Paris Observatory: SUPERSTRUCTURE \cite{Eis74}, DW \cite{Eis99}, JAJOM \cite{Sar72}, and AUTOLSJ \cite{TFR81a}. For our analysis we also used the Atomic Data and Analysis Structure (ADAS) data bank (conventionally called here ADAS set) \cite{ADAS}.

As was mentioned above (see Section 2.1), atomic data comprise atomic constants and collisional characteristics. Atomic constants include wavelengths, radiative and autoionization (for autoionizing states) decay probabilities $A^{(r)}(\gamma,\gamma')$ and $A^{(a)}(\gamma,\alpha_0)$ for transitions $\gamma\to\gamma'$ and $\gamma\to\alpha_0$, respectively, as well as the total transition probabilities $A^{(r)}(\gamma)=\sum_{\gamma''} A^{(r)}(\gamma,\gamma'')$, $A^{(a)}(\gamma)=\sum_{\alpha'} A^{(a)}(\gamma,\alpha')$. Using these data, the factors $F_2(\gamma)$ can be also determined needed for DS intensities:
\begin{equation}
F_2(\gamma) = \frac{g_{\gamma}}{g_0} \frac{A^{(a)}(\gamma,\alpha_0)\, A^{(r)}(\gamma,\gamma')}{A^{(a)}(\gamma) + A^{(r)}(\gamma)}    \, ,
\label{parameters}
\end{equation}
where $g_{\gamma}$ and $g_0$ are the statistical weights of the autoionizing state $\gamma$ and the ground state $\alpha_0$ of the recombining ion, respectively. Collisional characteristics of elementary processes involve cross-sections and rates for direct (potential or background) processes of electron-ion impact excitation, ionization, and radiative recombination. The contribution of resonance scattering was taken into account as a cascade process from autoionizing levels caused by dielectronic capture with following autoionization (resonance excitation) or radiative decay (dielectronic recombination). Effective rate coefficients for excitation and recombination processes including radiative cascades from upper excited and autoionizing levels were obtained by means of the collisional-radiative model (see \cite{GorThesis} for details of calculations). Radiative transition probabilities for {\bf x} and {\bf z} lines were taken from \cite{Joh95}.

All atomic characteristics within the framework of the PO set were calculated with the multi-configurational wave-functions. These functions were constructed from orbitals obtained in scaled Thomas-Fermi-Dirac-Amaldi potentials different for each {\it l} orbital. The scaling parameters were determined by minimization of the sum of the energies of all the terms belonging to the lowest configurations: $1snl$ ({\it n} = 1--4) for He-like ion states and $1s2lnl'$ ({\it n} = 2--5) for autoionizing states of Li-like ion. For autoionizing states with $n\ge 6$, the corresponding data were extrapolated from $n=5$ in a way similar to the papers \cite{TFR85} and \cite{Dubau85}. The free wave-function $\Psi_{f}^{i}(E)$ with one electron in the continuum corresponding to an autoionization channel were calculated with the same statistical potential to insure orthogonality between free and bound orbitals. Some relativistic corrections were accounted for by the intermediate coupling coefficients in the frame of a Breit-Pauli Hamiltonian (see \cite{GorThesis} for details of calculations).

The LPI data were obtained by means of perturbation theory expansion. Energy levels and probabilities of radiative and autoionization (for dielectronic satellites) decay were calculated by the MZ code using a {\it Z}-expansion method with a basis of hydrogen-like wavefunctions. Both electrostatic and relativistic (in the frame of the Breit Hamiltonian) interactions were treated as perturbation. Matrix elements of corresponding operators for degenerate states including configurations with the same parity and set of principle quantum numbers (Layzer complex) were expanded over $Z^{-1}$. The nonrelativistic energy matrix was expanded up to the second order. For radiative probabilities first order corrections over electrostatic interactions for wave-functions were taken into account. The improved data for autoionization probabilities accounting for the screening effect were obtained by means of the MZ code with radial integrals calculated by means of the ATOM code \cite{Gor06a}. Collisional characteristics (for direct processes) were calculated in the Coulomb-Born-exchange approximation modified by orthogonalizing free-bound wavefunctions for the exchange amplitude by means of the ATOM. In this code the radial orbitals for bound electrons were obtained in the effective scaled central potential with a scaling factor derived as an eigenvalue of the radial equation with a given energy. The configuration mixing coefficients were calculated by means of the MZ code (see \cite{Urn07,Gor03,GorThesis} for details of calculations).

The ADAS set of data used in this study is obtained on the base of the R-matrix calculations \cite{Whi02} and AUTOSTRUCTURE code \cite{Auto97}. Both codes are the major part of the ADAS project \cite{ADAS}. The AUTOSTRUCTURE, being the extension of SUPERSTRUCTURE, was used for the atomic structure calculations in both He- and Li-like systems of argon. Primarily it was utilized to generate the radial wave functions, necessary for the R-matrix calculations. The R-matrix data, providing the effective collisional strength up to $n=4$ in He-like ion were used as input parameters to calculate the excitation rates of the major spectral lines. For this purpose the radiative transition probabilities from AUTOSTRUCTURE were used. The calculations of the dielectronic satellites emitted from the Li- and Be-like ions were done using Slater type potential up to the states with $n=7$. These calculations demonstrated also the effect of the radiative transitions among the doubly excited states and their influence on the intensity of the satellites \textbf{q} and \textbf{r} \cite{Mar04}.

\begin{figure}
\resizebox{0.5\textwidth}{!}{%
  \includegraphics[angle=270]{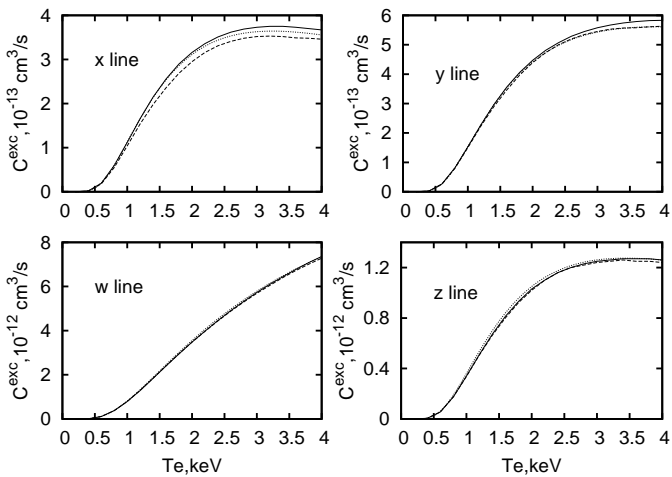}
}
\caption{Effective excitation rate coefficients calculated by means of ATOM (solid
lines), DW (dashed lines), and R-matrix (point lines) codes.}
\label{fig:2}       
\end{figure}

\begin{figure}
\resizebox{0.5\textwidth}{!}{%
  \includegraphics[angle=270]{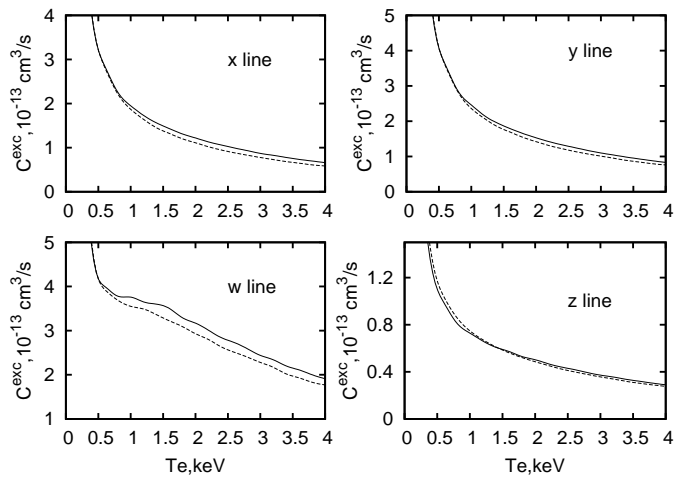}
}
\caption{Effective recombination rate coefficients calculated by means of ATOM (solid
lines) and DW (dashed lines) codes.}
\label{fig:3}       
\end{figure}

A comparison of the effective (including cascades) rate coefficients for collisional excitation of He-like argon ion lines as well as a few $F_2(\gamma)$ factors by LPI and ADAS were earlier given in \cite{Mar06a}. A comparison of the effective rate coefficients for collisional excitation by ATOM, DW, and ADAS is shown in Figure~\ref{fig:2}. In Figure~\ref{fig:3} the effective rate coefficients for the total (radiative and dielectronic) recombination by ATOM and DW are shown. Some of $F_2(\gamma)$ factors for transitions $1s2l2l' - 1s^2 2l''$ (column ``Key'' corresponds to the notations for satellite lines from \cite{Gab72}) in Li-like lines are shown in Table 1. For comparison we presented also the published results of calculations carried out by other methods (see also \cite{Gor06a}). The results of calculations in respect to the verification problem are discussed in Section \ref{sec:5}.

\begin{table*}
\begin{small}
\caption{Comparison of the $F_{2}(\gamma)$ factors for Ar$^{15+}$ dielectronic satellites.}
\begin{tabular}{cccccccccc}
\hline
Transition & Key & $\lambda $(\AA ) &  LPI & PO & ADAS & Pres. work & Ref. \cite{Chen86} & Ref. \cite{Nilsen88} \\
\hline
1s2p$^{2}$($^{2}$P$_{3/2}$) -- 1s$^{2}$2p($^{2}$P$_{3/2}$) & $\mathbf{a}$ & 3.9858 & 3.636  & 3.43  & 3.63 & 3.33  & 3.48  & 3.81  \\
1s2p$^{2}$($^{2}$P$_{3/2}$) -- 1s$^{2}$2p($^{2}$P$_{1/2}$) & $\mathbf{b}$ & 3.9818 & 0.258 & 0.264 & 0.276 & 0.236 & 0.242 & 0.268 \\
1s2p$^{2}$($^{4}$P$_{5/2}$) -- 1s$^{2}$2p($^{2}$P$_{3/2}$) & $\mathbf{e}$ & 4.0126 & 0.361 & 0.357 & 0.318 & 0.352 & 0.307 & 0.333 \\
1s2p$^{2}$($^{2}$D$_{5/2}$) -- 1s$^{2}$2p($^{2}$P$_{3/2}$) & $\mathbf{j}$ & 3.9939 & 22.87  & 23.0  & 22.8  & 22.3  & 21.5  & 23.2  \\
1s2p$^{2}$($^{2}$D$_{3/2}$) -- 1s$^{2}$2p($^{2}$P$_{1/2}$) & $\mathbf{k}$ & 3.9899 & 16.69  & 16.7  & 16.7  & 16.2  & 15.7  & 17.0  \\
1s2p$^{2}$($^{2}$S$_{1/2}$) -- 1s$^{2}$2p($^{2}$P$_{3/2}$) & $\mathbf{m}$ & 3.9656 & 2.28  & 2.67  & 2.38  & 2.10  & 2.60  & 2.55  \\
1s2p$^{2}$($^{2}$S$_{1/2}$) -- 1s$^{2}$2p($^{2}$P$_{1/2}$) & $\mathbf{n}$ & 3.9616 & 0.491 & 0.555 & 0.556 & 0.453 & 0.591 & 0.571 \\
1s2s2p($^{2}$P$_{3/2}$) -- 1s$^{2}$2s($^{2}$S$_{1/2}$) & $\mathbf{q}$     & 3.9815 & 0.732 & 1.42  & 1.25  & 0.645 & 2.07  & 1.12  \\
1s2s2p($^{2}$P$_{1/2}$) -- 1s$^{2}$2s($^{2}$S$_{1/2}$) & $\mathbf{r}$     & 3.9835 & 2.233  & 2.97  & 2.68  & 2.06  & 3.33  & 2.64  \\
1s2s2p($^{2}$P$_{3/2}$) -- 1s$^{2}$2s($^{2}$S$_{1/2}$) & $\mathbf{s}$    & 3.9677 & 1.649  & 2.07  & 2.97  & 1.64  & 2.42  & 1.82  \\
1s2s2p($^{2}$P$_{1/2}$) -- 1s$^{2}$2s($^{2}$S$_{1/2}$) & $\mathbf{t}$    & 3.9686 & 3.085  & 3.43  & 3.75  & 3.05  & 3.14  & 3.10  \\
\hline
\end{tabular}
\end{small}
\end{table*}

The recombination rates due to charge exchange in the process of collisions of highly ionized argon atoms with neutral atoms of the main gas (H) were calculated in \cite{Mar06c} by means of an analytical formula for cross-sections at low energies, $E{\leq}4$ keV \cite{Jan93}; for high energies, $E{\geq}15$ keV the values obtained in \cite{Hung01} were used. Corresponding rates for this process were obtained by averaging over the Maxwellian distribution with the ion temperature $T_{i}$.

\section{Experiment}
\label{sec:3}

TEXTOR is a medium-sized tokamak experiment with a major radius of 1.75 m and a minor radius of 0.46 m. It operates with toroidal magnetic fields of up to 2.7 T and plasma currents up to 580 kA. In addition to ohmic heating of about 0.5 MW, which is obtained from the plasma current, auxiliary heating is provided by the injection of neutral hydrogen beams with the total power of up to 2 MW. Further information on TEXTOR and its diagnostic equipment can be found in \cite{Ber99,Wei01,Ber04}. A series of discharges was performed for a wide range of temperatures, $T_e=0.8-2.5$~keV, and densities, $N_e=10^{13}-10^{14}$~cm$^{-3}$. The experiments were carried out in plasmas with hydrogen, deuterium, and helium atoms as the working gas. Additionally, two neutral hydrogen beams were injected into the plasma for heating, and also to investigate the effects of charge exchange of neutral hydrogen with argon ions on the spectra.

\begin{figure}
\resizebox{0.5\textwidth}{!}{%
  \includegraphics{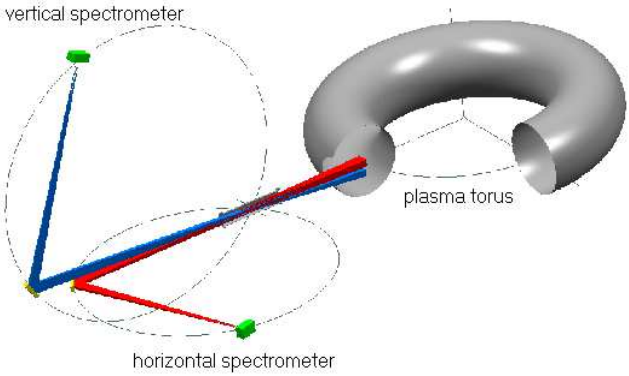}
}
\caption{Scheme of the X-ray spectrometer installed at the TEXTOR tokamak (by courtesy of G.~Bertschinger).}
\label{fig:4}
\end{figure}

The {\it K}-spectra from Ar$^{16+}$ and Ar$^{15+}$ impurity ions were taken by means of a high resolution X-ray spectrometer/polarimeter \cite{Ber99,Wei01} consisting of two (horizontal and vertical) Bragg spectrometers in the Johann scheme, each employing cylindrically bent $153\,\mathrm{mm}\times 38\,\mathrm{mm}\times 0.7\,\mathrm{mm}$ quartz 110 crystal with a 2d spacing of 4.913 \AA\ and a 1-D position-sensitive detector. These spectrometers were designed for the polarization measurements of the radiation from the same central region of the tokamak plasma. In our studies it was also used the experimental scheme where the horizontal instrument was used to measure the $K_{\alpha}$-spectra (formed by transitions $n=2\to n=1$ of the optical electron), while the vertical (perpendicularly arranged) spectrometer was used in the experiment for recording the $K_{\beta}$-emission lines (emitted due to transitions $n=3\to n=1$). Figure~\ref{fig:4} shows a schematic diagram of the experimental setup.

The radius of curvature of the crystals was 3820~mm for the horizontal spectrometer and 4630~mm for the vertical one. Each detector consisted of a multi-wire proportional counter from the former X-ray crystal spectrometers at the Tokamak Fusion Test Reactor in Princeton. It has a large entrance window of 180~mm$\times$90~mm, a high count rate capability of up to 2.5$\times$10$^{5}$ photons/s, and a spatial resolution of 0.4 to 0.65~mm, depending on the count rate. Optimizations of the polarimeter resulted in a 70~mm$\times$8.5~mm exposed area for the horizontal crystal and a 30~mm$\times$13~mm exposed area for the vertical one. The resulting spectral resolution for the horizontal system was $\lambda /\delta\lambda $ = 5600 when the detector is optimized for count rate and $\lambda /\delta\lambda $ = 8300 when the detector was optimized for resolution corresponding to ion temperatures of 0.22 keV and 0.10 keV, respectively. For the vertical system the spectral resolution was $\lambda /\delta\lambda $ = 7200 in the former case and $\lambda /\delta\lambda $ = 10300 when the detector was optimized for resolution, and corresponding ion temperatures were 0.13 keV and 0.07 keV, respectively. The wavelength precision was estimated to 10$^{-4}$\AA . The data acquisition system have made it possible to record up to 8192 spectra per discharge and to cover the full discharge with a time resolution of 0.5 ms per spectrum.

The central line of sight of the spectrometer was at the angle $\mathbf{\alpha }_{0}$ = 10$^\circ$ with respect to the major radius of the plasma torus to make it possible to observe a component of about 17\% of the toroidal plasma velocity from the Doppler shift
of the spectral lines. The value of $\mathbf{\alpha}$ varied from 11.2$^\circ$ for the line $\mathbf{w}$ to 9.0$^\circ$ for the line $\mathbf{z}$. The central Bragg angle of the spectrometer was set to 54$^\circ$.

The electron temperature and density profiles $P_{\nu}(\rho)$ for $\nu =\{T_{e},N_{e}\}$ were measured for each spectrum by means of an electron cyclotron emission (ECE) polychromator and a far-infrared interferometer/polarimeter (FIIP), respectively \cite{Wei01}. The measured profiles were parameterized using the least square method by the expression:
\begin{equation}
P_{\nu }(\rho )=\alpha+\left( 1-\alpha\right)\left( 1-\rho ^{\beta }\right)^{\gamma
} \, .  \label{radial profiles}
\end{equation}
The shift of the magnetic surfaces toward the low-field side of the tokamak, $\mathrm{\Delta}$, was also modeled using the similar relation and was obtained from the measured profile of plasma pressure ($\mathrm{\Delta}(\rho)=\mathrm{\Delta_0} \cdot[1-\rho^{\gamma}]^{\eta}$, where $\mathrm{\Delta_0}$, $\gamma$, and $\eta$ are the fitting parameters). Since $P$-profiles for ion temperature are not routinely measured, it was assumed that the normalized profiles $\mathbf{P}_{T_{i}}$ are equal to that of the electron temperature. This assumption is more
valid at the higher densities where coupling between ions and electrons is good. It is important to note that the ion temperature profile only affects the deduced ion temperature from the spectra and has little effect on the relative intensities.

The central ion temperature $T_{i}^{(0)}$ was obtained by a fit to the widths of the lines with the help of the fitting procedure (FP) described below. Voigt profiles were used to produce all spectral lines. The Lorentzian component was set according to the lifetime of the upper level for each transition, and a contribution from the rocking curve of the crystal was added. The Gaussian component was set by the instrumental and the Doppler widths. The detector intersected with the Rowland circle, since it is plane and is placed perpendicular to the direction of the incident X-rays. As a result, the system was completely in focus for only one wavelength. The spectrometers were set such that the $\mathbf{Z}$ peak was on the Rowland circle. To account for defocusing of the spectrometers along the direction of dispersion, ray tracing was employed. The rocking curve was assumed to be Lorentzian with a spectral resolution of $3\times 10^{4}$, corresponding to the resolution for a flat quartz 110 crystal. Voigt profiles were inputted to the ray tracing corresponding to the instrumental profile excluding the effects of the rocking curve. Ray tracing was performed for many positions on the detector to determine the effect of defocusing. The resulting profiles were fit to a Voigt function to determine the Lorentzian and Gaussian contributions to the defocusing of the lines along the detector. Defocusing was small enough so the line shape remained close to a Voigt profile.

\section{SCA constituents}
\label{sec:4}

\subsection{Main equations}
\label{subsec:41}
Modeling of the tokamak spectra is usually based on the balance equations for the population densities $N_{z}^{(k)}$ of impurity ions. In the case of cylindrical geometry these equations can be written in the form:
\begin{eqnarray}
\partial_{t}N_{z}^{(k)}+\frac{1}{r}\partial _{r}(r\Gamma_{z}^{(k)}) = \nonumber \\
= \sum_{z^{\prime },k^{\prime }\neq z,k}\left[
N_{z^{\prime }}^{(k^{\prime })}\ae (z^{\prime }k^{\prime
},zk)-N_{z}^{(k)}\ae (zk,z^{\prime }k^{\prime })\right] \, ,
\label{basic equations}
\end{eqnarray}
where the matrix elements $\ae (a,b)$ are probabilities in $[s^{-1}]$ of transitions $a\rightarrow b$ for elementary processes in the plasma and $\Gamma_{z}^{(k)}$ is the radial flux density of argon ions expressed as
\begin{equation}
\Gamma_{z}^{(k)}=[-D_{z}^{(k)}(r)\partial_{r}+v_{z}^{(k)}(r)]N_{z}^{(k)}.
\label{radial flux}
\end{equation}
Here $D_{z}^{(k)}(r)$ and $v_{z}^{(k)}(r)$ are diffusion coefficients and convection velocities, respectively. In the general case of \emph{ab initio} calculations of the spectra\ these equations should be complemented by a system of similar equations for the population densities of the working gas states and MHD equations for plasma characteristics; in the semi-empirical models restricted by Eqs. (\ref{basic equations}) the values, characterizing the plasma states (in particular, radial profiles for electron temperature and density), are just the variable
parameters which have to be derived from measurements.

At low electron densities ($N<N^{\ast}$, where $N^{\ast }$ is the critical density discussed below) coronal approximation for the population densities of excited states (coronal equilibrium for populations $N_{z}^{(k)}$) takes place, implying a quasi-steady-state of the plasma ($\partial_{t}N_{z}^{(k)}=0 $), when all pumping processes in Eqs. (\ref{basic equations}) are caused by collisions (\ae $^{(col)}$) and are balanced by spontaneous radiative and autoionization (Auger) decay (\ae $^{(dec)}$). In hot plasmas for the states $k>1$ the effects of ionic and radiative transfer are negligible ($\partial_{r}N_{z}^{(k)}=0$, and an optical thickness $\tau \ll 1)$. In this approximation the balance Eqs. (\ref{basic equations}) for excited states take the form
\begin{equation}
\sum_{z^{\prime },k^{\prime }\neq z,k}\left[ N_{z^{\prime
}}^{(k^{\prime })}\ae (z^{\prime }k^{\prime
},zk)-N_{z}^{(k)}\ae (zk,z^{\prime }k^{\prime })\right]
=0, \,\, k>1, \label{modelEqs}
\end{equation}
and the spectral emissivity $j(P_{\nu };\rho ;\lambda )$ and the emissivity function for the line $l$ (contribution function) $j_{l}(T_e,N_g,N_e;\rho)$ $\rightarrow$ $j_{l}(T_e,N_g;\rho)$ in Eq. (\ref{emiss}) are given by
\begin{equation}
j(P_{\nu };\rho ;\lambda )=\sum_{z}j^{z}(T_{e},N_{g};\lambda
)n_{z}(\rho )\, ,  \label{cor-spec-emiss}
\end{equation}

\begin{equation}
j_{l}(T_{e},N_{g};\rho )=\sum_{z}C_{l}^{z}(T_{e},N_{g})n_{z}(\rho
)\, , \label{coronal-l-emissivity}
\end{equation}

\begin{equation}
j^{z}(T_{e},N_{g};\lambda
)=\sum_{l}C_{l}^{z}(T_{e},N_{g})\, \pounds
_{l}(T_{e};\lambda -\lambda _{l})\, ,  \label{partial-rates}
\end{equation}
where $j^{z}(T_{e},N_{g};\lambda )$ and $C_{l}^{z}(T_{e},N_{g})$ are, respectively, the partial emissivities and the effective rates (cross-sections averaged over Maxwellian distribution function) corresponding to the processes of the line excitation from ions with the charge $z$ (including cascades from upper levels and the branching ratio $k_{l}$ for the line $l$).

In the coronal approximation the populations of the ground states of ions are equal to the ionic abundances $N_{z}^{1}=N_{z}$ and the ionic densities $N_{z}$ obey the balance Eqs. (\ref{basic equations}) which can be written as
\begin{eqnarray}
\frac{1}{r}\partial _{r}(r\Gamma _{z})=\sum_{z^{\prime }\neq
z,}\left[ N_{z^{\prime }}\ae (z^{\prime },z)-N_{z}\ae
(z,z^{\prime })\right] \, , \nonumber \\
\Gamma_{z}=[-D_{z}(r)\partial _{r}+v_{z}(r)]N_{z}\, ,
\label{ionEquilibrium}
\end{eqnarray}
where $\ae (z^{\prime },z)=N_{e}C_{z}^{z^{\prime }}(T_{e})$ and $C_{z}^{z^{\prime }}(T_{e})$ are the corresponding total rates including all channels of recombination ($z^{\prime }<z$) and ionization $(z^{\prime }>z)$ processes. In particular, as it was pointed out in \cite{Ros99,Mar06c}, the total recombination rate $C_{z}^{TR}(T_{e})=C_{z}^{z-1}(T_{e})$, besides the radiative recombination (RR) and dielectronic one (DR) also has to include the effect of charge transfer from neutral atoms of the main gas (CXR): $C_{z}^{TR}(T_{e})=C_{z}^{RR}(T_{e})+C_{z}^{DR}(T_{e})+n_{g}C_{z}^{CRX}(T_{i},E_{g}) $, where $n_{g}=N_{g}/N_{e}$ and $E_{g}$ is the energy of the neutral particles from the beam. In the case of the ohmically heated plasmas, however, the neutral particles are characterized by the temperature $T_{g}$, equal to the ion temperature $T_{i}$. It is worth noting here that at the electron densities typical for tokamak plasmas the (spatially non-equilibrium) ionic densities $N_{z}$ satisfying Eq. (\ref{ionEquilibrium}) are not equal to that $N_{z}^{C}$, obtained in the (coronal) approximation applied for conditions of the solar corona (so-called ``coronal ionization equilibrium'', CIE). The CIE is usually calculated under two additional assumptions for local equilibrium: neglecting the spatial derivative ($\partial _{r}=0$) as well as the CXR channel of recombination. To avoid confusion, in distinction to CIE densities $N_{z}^{C}$, we will denote through $N_{z}^{E}$ the equilibrium values ($\partial _{t}=\partial _{r}=0)$ derived from the balance equations (\ref{ionEquilibrium}) with account for the CXR.

The flux $F_{[\mathbf{L}]}$ and the emissivity function $J_{[\mathbf{L} ]}(T_{e},N_{g};\rho )$ for the peak in Eq. (\ref{peak-flux}) are expressed through the partial excitation rate $J_{[\mathbf{L} ]}^{z}(T_{e},N_{g})$, defined as
\begin{eqnarray}
J_{[\mathbf{L}]}^{z}(T_{e},N_{g})=\int_{[\mathbf{L}]}j^{z}(T_{e},N_{g};\lambda
)d\lambda =  \nonumber \\
=\sum_{l}C_{l}^{z}(T_{e})\pounds _{[\mathbf{L}]}^{l} \, ,
\label{Lpartial_exc-rate}
\end{eqnarray}
by the expressions

\begin{eqnarray}
F_{[\mathbf{L}]}=\sum_{z}F_{[\mathbf{L}]}^{z} =  \nonumber \\  = C\sum_{z}\int_{0}^{1}J_{[\mathbf{L}]}^{z}(T_{e},N_{g})n_{z}(\rho
)[\,N_{e}(\rho )]^{2}\,g(\rho )\,d\rho \, ,
\label{coronal-L-flux}
\end{eqnarray}

\begin{equation}
J_{[\mathbf{L}]}(T_{e},N_{g};\rho )=\sum_{z}J_{[\mathbf{L}
]}^{z}(T_{e},N_{g})n_{z}(\rho )\, , \label{peak-emiss-function}
\end{equation}
where $F_{[\mathbf{L}]}^{z}$ denotes the partial fluxes and $\pounds _{[\mathbf{L}]}^{l}$ are the correction factors due to the shape of lines $l$, contributing to the wavelength range of
the peak $\mathbf{L}$ (note here that the position of the line $l$ could be close to the confines or even outside of the range $[\mathbf{L}]$)
\begin{equation}
\pounds
_{[\mathbf{L}]}^{l}(T_{i})=\int_{[\mathbf{L}]}\pounds_{l}(T_{i};\lambda
-\lambda _{l})d\lambda . \label{shape-correction}
\end{equation}
The value $J_{[\mathbf{L}]}^{z}$ describes the partial effective rate of formation $C_{l}^{z}$ for all lines, contributing to the peak $\mathbf{L}$, from ions with charge $z$. Thus for $z$=16 (denoted also as $he$) the partial emissivity $J_{[\mathbf{L}]}^{he}$ contains the effective excitation rates for He-like ions lines and the rates of dielectronic capture for
Li-like dielectronic satellites arising from He-like ions; for the lines excited from Li-like ions ($z$=15 or $li$) $J_{[\mathbf{L}]}^{li}$ includes the effective rate for the inner-shell excitation of Li-like satellites and the inner-shell ionization (for excitation of the $z$ line); $J_{[\mathbf{L} ]}^{h}$ ($\mathbf{z}$=17 or $h$) contains the total rate of
recombination of H-like ions to the excited states of He-like lines.

For the following analysis it is important to introduce the dimensionless factors $G_{\mathbf{L}}$ defined as:
\begin{equation}
G_{\mathbf{L}}(P,\mathbf{P})=\frac{F_{[\mathbf{L}]}(P,\mathbf{P})}
{F_{[\mathbf{L}]}^{c}(P)}\,.  \label{G-L-value}
\end{equation}
Here $F_{[\mathbf{L}]}^{c}(P)$ is the flux intensity from the core of the tokamak plasma in the ``core approximation'':
\begin{eqnarray}
F_{[\mathbf{L}]}^{c}(P) = C\, Y\, J_{[\mathbf{L}]}(T_{e0})n_{z}(T_{e}^{(0)})\, , \label{central-flux} \\
Y=\int_{0}^{1}[N(\rho )]^{2}\,g(\rho )d\rho \, . \nonumber
\end{eqnarray}
These factors provide the measure for the deviation of the flux in the synthetic spectrum from that in the core approximation caused by the $P_{\nu }$-profiles. For two peaks $\mathbf{L}_{1}$ and $\mathbf{L}_{2}$ it is also convenient to introduce the relative
value for G-coefficients
\begin{equation}
G_{\mathbf{L}_{2}}^{\mathbf{L}_{1}}=G_{\mathbf{L}_{1}}/G_{\mathbf{L}_{2}}\, ,
\label{G-L1-L2-factor}
\end{equation}
which characterizes the deviation from the core approximation for the ratio of fluxes in these peaks.

\subsection{Fitting procedure}
\label{subsec:42}
Assuming that all normalized radial profiles $\mathbf{P}(\rho )$ are known (i.e., fixed), the synthetic spectra are governed by the central plasma core parameters included in the set $P=\{P_{\nu }\}$. For application of the fitting procedure (FP), it is convenient to introduce the dimensionless variable fitting parameters $f_{\nu }$ defined as follows: $f_{e}=T_{e}^{(0)}/T_{m}$, $f_{i}=T_{i}^{(0)}/T_{m}$, $f_{li}=\widetilde{n}_{li}(0)$, and $f_{h}=\widetilde{n}_{h}(0)$, where $T_{m}$ is the maximum of the electron temperature in the measured radial profile and $\widetilde{n}_{z}(\rho )$ for $z=\{li,h\}$ are the relative (to density of He-like ions) radial profiles for ionic abundances $\widetilde{n}_{z}$=$n_{z}/n_{he}$. The synthetic spectrum (\ref{spectral intensity}) written by means of (\ref{cor-spec-emiss}) in the form
\begin{eqnarray}
I(\lambda )=C\int_{0}^{1}( j^{he}(\rho ;\lambda )+j^{li}(\rho ;\lambda )f_{li}\widetilde{n}_{li}(\rho )+  \nonumber \\
+ j^{h}(\rho ;\lambda )\,f_{h}\widetilde{n}_{h}(\rho ) ) n_{he}(\rho
)\left( N_{e}(\rho )\right)^{2} d\rho  \label{synthetic}
\end{eqnarray}
was used to obtain three plasma parameters $f_{e}$, $f_{li}$, and $f_{h}$ by iterative fitting the measured spectra in the spectral ranges $\mathbf{[L]}$ minimizing the $\chi^{2}$ values
\begin{equation}
\chi ^{2}=\int_{[\mathbf{L}]}\left\vert \frac{I^{\exp }(\lambda )-I(\lambda )}{I^{\exp }(\lambda
)}\right\vert ^{2} d\lambda \, . \label{chi}
\end{equation}

The peaks for the determination of the plasma parameters ($\mathbf{W,X,Y,Z,K}$) were chosen such as to provide the best statistics of the flux and the maximum sensitivity of the peak intensity on the corresponding parameters as well as to minimize the number of key parameters stipulating the flux ratio. In particular, the total flux in the triplet lines does not depend on the mixing coefficients between \textbf{x} and \textbf{z} lines and on the overlapping of \textbf{x} and \textbf{y} line shapes. Other peaks were used to test the accuracy of atomic data and the central temperature diagnostics.

The constant $C$ was defined for the range $[\sigma ]$ with $\sigma = \{\mathbf{W}\}$. The experimental spectra $I^{\exp}(\lambda)$ in Eq. (\ref{chi}) were derived accounting for the contribution of the background emission; the background on the detectors was fitted using a parabolic function. The detailed description of the fitting procedure as well as its application for plasma diagnostic purposes is given in \cite{Wei01} and \cite{Mar06a,Mar06b}; below we will present only a short description of how it was applied to the problem of atomic data verification.

\subsection{Spectroscopic model}
\label{subsec:43}
The aforementioned SM is based on the following assumptions:

\smallskip
\noindent 1. The conditions necessary for coronal equilibrium are satisfied for populations of excited states ($k>1$).

\noindent 2. The profiles of relative abundances for argon ions $n_{z}(\rho )$ satisfy the equation
(\ref{continuity eq}); normalized profiles of electron temperature $\mathbf{T}_{e}$ and density $\mathbf{N}_{e}$ are known.
\smallskip

In distinction to \emph{ab initio} models, employing the balance equations (\ref{basic equations}) for all states of impurity ions, in the SM these equations are used only for excited ones and hence the relative ionic abundances $n_{z}(\rho )$ are considered as the
variable parameters of the model.

To formulate the SM and to identify its key parameters, it is necessary to factorize the dependence on radial profiles of plasma parameters $n_{z}(\rho )$, $N_{e}(\rho )$, $T_{e}(\rho )$ in the main equation for the flux intensity in the peak (\ref{coronal-L-flux}), changing the variable of integration $\rho $ to the dimensionless temperature variable $\beta (\rho
)=[\mathbf{T}_{e}(\rho )]^{-1}-1$:
\begin{equation}
F_{[\mathbf{L}]}=C\sum_{z}\int_{0}^{\beta _{1}}J_{[\mathbf{L}
]}^{z}(T_{e}^{(0)};\beta )\,n_{z}(T_{e}^{(0)},\beta )\,y(\beta )g(\beta
)\,\,d\beta , \label{L-flux-betha-represent}
\end{equation}
where $y(\beta )$=$[N(\rho (\beta ))]^{2} \left|d\rho / d\beta \right|$ is the {\it differential emission measure} (DEM), and $\beta _{1}$=$\beta (1)\gg$1. In the ``$\beta$-representation'' in distinction to ``$\rho$-representation'' in Eq. (\ref{coronal-L-flux}) the values $J_{[\mathbf{L}]}^{z}$ depend only on central values of
plasma parameters (to simplify notations below we will omit the dependence on $T_{i}^{(0)}$ and $n_{g}(0)$ accounting for in calculations; note here that these parameters, though important
for diagnostics of plasmas characteristics, practically does not influence the results of atomic data verification).

For application of the Bayesian iterative method (BIM, see below Section 4.4) to inversion procedure, it is also necessary, besides the total $F_{[\mathbf{L}]}$ and partial $F_{[\mathbf{L}]}^{z}$ fluxes for the peaks $\mathbf{L}$, to introduce the relative fluxes $\Gamma_{[\mathbf{L}]}(\sigma _{z})=F_{[\mathbf{L} ]}/F_{[\mathbf{\sigma }_{z}]}$ and $\Gamma_{[\mathbf{L}]}^{z}=F_{[\mathbf{L} ]}^{z}/F_{[\mathbf{\sigma }_{z}]}$, normalized in spectral regions $\mathbf{[\sigma }_{z}]$, chosen for each $z$. To identify these
regions consider three sets of peaks denoted through $\mathbf{Z}=\{\mathbf{L}\}$ ($\mathbf{Z=Li,He,H})$, corresponding to three (overlapped) wavelength regions $[\mathbf{\sigma }_{z}]=[\mathbf{Z]},$ which include lines containing partial excitation from ions with the charge $z=\mathbf{Z}$ ($z=li,he,h$). The choice of these sets was stipulated by two demands: in each region $[\mathbf{Z}]$ to provide maximum contribution of partial fluxes $F_{[\mathbf{L}]}^{z}(\mathbf{L\subset Z)}$, having different dependence on $\beta$, and minimum contribution of partial fluxes $F_{[\mathbf{L}]}^{k}$ from ions with charge $k\neq z$.

In the spectral range [$\mathbf{Z}$] the total flux $F_{[\mathbf{Z}]}$ can be written then as a sum of partial fluxes $F_{[\mathbf{Z}]}^{z}$ similar to that in the range [$\mathbf{L}]$ (see Eq. (\ref{coronal-L-flux})):
\begin{equation}
F_{[\mathbf{Z}]}=\sum_{z}F_{[\mathbf{Z}]}^{z}
\label{partial-Z-fluxes}
\end{equation}
and the partial luminosity function $J_{[\mathbf{Z}]}^{z}$ as a sum over all $\mathbf{L}\subset \mathbf{Z}$:
\begin{equation}
J_{[\mathbf{Z}]}^{z}(T_{e}^{(0)},\beta )=\sum_{\mathbf{L\subset
Z}}J_{[\mathbf{L}]}^{z}(T_{e}^{(0)},\beta )
\label{J-partial}
\end{equation}
Introducing three normalized functions $\Phi _{z}(T_{e}^{(0)},\beta )$ and the normalized partial excitation rates $p_{[\mathbf{L}]}^{z}$ ($\beta$-profiles for $z$=15--17) defined as
\begin{equation}
\Phi _{z}(T_{e}^{(0)},\beta ) =\frac{C J_{[\mathbf{Z}]}^{z}(T_{e}^{(0)},\beta
)n_{z}(T_{e}^{(0)},\beta )\,y(\beta )g(\beta )}{F_{[\mathbf{Z}]}^{z}}
\label{z-profiles}
\end{equation}
and
\begin{equation}
p_{[\mathbf{L}]}^{z}(T_{e}^{(0)},\beta
)=J_{[\mathbf{L}]}^{z}(T_{e}^{(0)},\beta
)/J_{[\mathbf{Z}]}^{z}(T_{e}^{(0)},\beta )\, ,
\label{i-p-profiles}
\end{equation}
the relative flux $\Gamma _{[\mathbf{L}]}$ for the peak $\mathbf{L\subset Z}$ in the range [$\mathbf{Z}$] can then be written in the form
\begin{eqnarray}
\Gamma_{[\mathbf{L}]}(T_{e}^{(0)}) &=& \sum_{z=\mathbf{K}}\Gamma_{[\mathbf{K}]}^{z}(T_{e}^{(0)})R_{\mathbf{Z}}^{\mathbf{K}}P_{[\mathbf{L}
]}^{z}(T_{e}^{(0)}) \,\,\mathrm{for}\,\, \mathbf{L\subset Z} \, , \nonumber  \\ \Gamma_{[\mathbf{Z}]} &=& \sum_{\mathbf{L\subset
Z}}\Gamma_{[\mathbf{L}]}=1\, ,  \label{gamma}
\end{eqnarray}
where $R_{\mathbf{Z}}^{\mathbf{K}}=F_{[\mathbf{K}]}/F_{[\mathbf{Z}]}$, $\Gamma_{[\mathbf{Z}]}^{z}=F_{[\mathbf{Z}]}^{z}/F_{[\mathbf{Z}]}$ and the partial flux in the peak $[\mathbf{L}]$, $F_{[\mathbf{L}]}^{z}$, normalized on that in the range $[\mathbf{Z}]$,
$P_{[\mathbf{L}]}^{z}=F_{[\mathbf{L}]}^{z}/F_{[\mathbf{Z}]}^z$, is expressed through partial excitation rates $p_{[\mathbf{L}]}^{z}$ and profiles $\Phi _{z}(T_{e}^{(0)},\beta )$ by the equalities:
\begin{equation}
P_{[\mathbf{L}]}^{z}(T_{e}^{(0)})=\int_{0}^{\beta _{1}}p_{[\mathbf{L}]
}^{z}(T_{e}^{(0)},\beta )\Phi _{z}(T_{e}^{(0)},\beta )\,\,d\beta \, .
\label{flux}
\end{equation}
These values satisfy the normalization conditions following from their definitions and Eqs. (\ref{gamma}):
\begin{eqnarray}
\sum_{\mathbf{L\subset
Z}}P_{[\mathbf{L}]}^{z}(T_{e}^{(0)})=\sum_{\mathbf{L\subset Z}} p_{[\mathbf{L}]}^{z}(T_{e}^{(0)})=1\, ;\nonumber \\ \int_{0}^{\beta
_{1}}\Phi _{z}(T_{e}^{(0)},\beta )\,\,d\beta =1\, .
\label{normalization-1}
\end{eqnarray}

The relative partial fluxes $\Gamma _{[\mathbf{Z}]}^{z}$ in three spectral ranges $[\mathbf{Z}]$ ($\mathbf{Z=Li,He,H}$) can be expressed through the flux ratios $R_{\mathbf{Z}}^{K}$ and the normalized integrals $P_{[\mathbf{Z}]}^{k}(T_{e}^{(0)})$, obtained by solving the set of Eqs. (\ref{gamma})
\begin{equation}
1=\sum_{k=\mathbf{K}}\Gamma
_{[\mathbf{K}]}^{k}(T_{e}^{(0)})R_{\mathbf{Z}
}^{\mathbf{K}}P_{[\mathbf{Z}]}^{k}(T_{e}^{(0)})\,\, \mathrm{for}\,\, \mathbf{Z=Li,He,H}\, ,   \label{partial-gamma}
\end{equation}
providing the close set of equations (\ref{gamma}), (\ref{flux}), and (\ref{partial-gamma}) referred to as the SM.

The flux intensities expressed by the these equations are thus stipulated by three key plasma parameters $\Phi _{z}(T_{e}^{(0)},\beta)$ ($\Phi_z$-profiles) as well as the kernels of the integral equations (\ref{flux}) $p_{[\mathbf{L]}}^{z}(T_{e}^{(0)},\beta )$. These kernels besides the set of atomic data {\bf D} depend on the center temperature of the plasma core $T_{e}^{(0)}$ and (for $z=h$ and $Z\subset He)$ on $N_{g}$-profile. At the condition (\ref{exp=th}), $F_{[\mathbf{L}]}^{\exp }=\sum_{z}P_{[\mathbf{L}]}^{z}(T_{e}^{(0)})$ each $\Phi _{z}$-profile
(for $z$ = $li,he,h)$, as is seen from Eq. (\ref{flux}), depends only on those parameters which are the arguments of $p_{[\mathbf{L]}}^{z}$.

If all relative fluxes in the peaks $\Gamma _{\lbrack \mathbf{L}]}$ and the set of atomic data are known (fixed), the $\Phi _{z}(T_{e}^{(0)},\beta )$ functions can be found solving the inverse problem for a set of model equations at any given values of the $T_{e}$ and $N_{g}$-profile. Using the definition of the $\Phi _{z}$-profiles (\ref{z-profiles}), the abundances
$n_{z}(T_{e}^{(0)},\beta )$ and their sum $n(T_{e}^{(0)},\beta )$ for $z=li,he,h$ can be expressed through $\Phi _{z}(T_{e}^{(0)},\beta )$ and $n_{he}(T_{e}^{(0)},\beta )$ as
\begin{eqnarray}
n_{z}(T_{e}^{(0)},\beta ) &=& \widetilde{n}_{z}(T_{e}^{(0)},\beta
)\,n_{he}(T_{e}^{(0)},\beta )\, , \nonumber  \\
n(T_{e}^{(0)},\beta ) &=& n_{he}(T_{e}^{(0)},\beta )\sum_{z}\widetilde{n}_{z}(T_{e}^{(0)},\beta )\, ,
\label{abundances}
\end{eqnarray}
where
\begin{equation}
\widetilde{n}_{z}(T_{e}^{(0)},\beta ) = \frac{\Phi _{z} }{\Phi
_{he}}\frac{J_{[\mathbf{He}]}^{he}}{J_{[\mathbf{Z}]}^{z}}
\frac{\Gamma_{[\mathbf{Z}]}^{z}}{\Gamma_{[\mathbf{He}]}^{he}}
R_{\mathbf{He}}^{\mathbf{Z}}\, .
\label{rel_abund}
\end{equation}
Applying the SM for diagnostic purposes the relative (to He-like ion density) ionic abundances $\widetilde{n}_{z}(T_{e}^{(0)},\beta )$ for $z=h,li$ may be then directly derived from Eq. (\ref{rel_abund}); with the help of the additional normalization condition (\ref{continuity eq}) one may also obtain the relative abundances $n_{z}(T_{e}^{(0)},\beta )$ for all three ions.
On the other hand, this equality followed from physical requirements restricts the class of possible formal solutions of the inverse problem \{$\Phi _{z}^{Inv}(T_{e}^{(0)},\beta )\}$ and hence can be also used for the following optimization of the central temperature $T_{e}^{(0)}$ and the correction of the key atomic parameters from the set $\mathbf{D}$. However, since it is
expressed in terms of the relative ionic abundances $n_{z}(T_{e}^{(0)},\beta )$, rather than the $\Phi _{z}(T_{e}^{(0)},\beta )$ functions, one has to adopt an additional (experimental in SM)
information on plasma parameters, namely the profile of the DEM $y(\beta )$, depending on the temperature and density profiles $P_{Te}(\beta )$ and $P_{Ne}(\beta )$, respectively.

Using the identity $n(T_{e}^{(0)},\beta )=n(T_{e}^{(0)},\beta )/n(T_{e}^{(0)},\beta _{c})$ for some fixed values of $\beta =\beta_{c}$ within the interval $\beta _{\min }\leq \beta _{c}\leq \beta
_{\max }$, where Eq. (\ref{continuity eq}) is satisfied ($n(T_{e}^{(0)},\beta _{c})=1)$, one obtains the condition for $n(T_{e}^{(0)},\beta )$ at any $\beta$
\begin{eqnarray}
n(T_{e}^{(0)},\beta )=\frac{y(\beta _{c})}{y(\beta )}\left[
\sum_{z}\frac{\Phi _{z}(T_{e}^{(0)},\beta )}{J_{[Z]}^{z}(T_{e}^{(0)},\beta
)}\frac{\Gamma_{[Z]}^{z}}{\Gamma_{[He]}^{he}}R_{He}^{Z} \right]\times \nonumber  \\
\times \left[ \sum_{z} \frac{\Phi _{z}(T_{e}^{(0)},\beta
_{c})}{J_{[Z]}^{z}(T_{e}^{(0)},\beta
_{c})}\frac{\Gamma_{[Z]}^{z}}{\Gamma_{[He]}^{he}}R_{He}^{Z}\right]^{-1}=1\, ,
\label{sum}
\end{eqnarray}
which is convenient to use for the optimization of parameters $T_{e}^{(0)}$ and $\mathbf{D}$ when the DEM $y(\beta )$ is known. The boundaries of the interval ($\beta _{\min },\beta _{\max })$ are connected with the contribution of nuclear (at the values of $\beta \leq $ $\beta _{\min }$) and low $Z$ ions (Be-like, B-like, etc. at $\beta \geq \beta _{\max }$) which were not accounted for in the sum $n(T_{e}^{(0)},\beta )$ (note that the deviation of $n(T_{e}^{(0)},\beta )$ from unity for small and large $\beta$ makes it possible to estimate the abundances of these ions for high and low temperature spectra).

It should be mentioned here that the calculations carried out in the present work by means of a collisional-radiative model for the argon ions showed a deviation from the coronal approximation of about 1\% for the levels with $n=2$ at the electron density 10$^{14}$~cm$^{-3}$, indicating the validity of the assumptions for the SM at the densities $N_{e}<N^{\ast }=$~10$^{14}$~cm$^{-3}$.

\subsection{BIM inversion}
\label{subsec:44}
To solve the inverse problem for deriving $\Phi_{z}$-profiles, we have used the Bayesian iterative method (BIM) which was previously applied as an effective diagnostic tool for analyzing and interpreting XUV spectral data from hot tokamak and solar corona plasmas \cite{Urn07,Zhi87,Zhi06,Urn07b,Gor10,Urn12}.

Defining the values $P_{[\mathbf{L}]}^{z}(\exp )$ and $\Gamma_{\lbrack \mathbf{L]}}^{z}(\exp )$ by means of the expressions
\begin{eqnarray}
P_{[\mathbf{L}]}^{z}(\exp ) &=& \Gamma_{[\mathbf{L}]}^{z}(\exp) / \Gamma_{[\mathbf{Z}]}^{z}\, ,  \nonumber  \\
\Gamma_{[\mathbf{L}]}^{z}(\exp ) &=& \Gamma_{[\mathbf{L}]}^{\exp
}-\sum\limits_{k\neq z}\Gamma_{[\mathbf{L}]}^{k}\, , \,\, \mathrm{for}\,\, \mathbf{L\subset [Z]}\, ,  \label{P(exp)}
\end{eqnarray}
where $\Gamma_{[\mathbf{L]}}^{\exp }$ stands for the measured flux ratios of the peaks, and using the relation (\ref{gamma}) and the definition of the $P_{[\mathbf{L}]}^{z}$ in (\ref{flux}) at the condition $P_{[\mathbf{L}]}^{z}(\exp )=P_{[\mathbf{L}]}^{z}(T_{e}^{(0)})$, one can obtain the system of self-consistent equations for {\bf Z}=$z$=15--17 and $\mathbf{L}\subset\mathbf{Z}$:
\begin{equation}
P_{[\mathbf{L}]}^{z}(\exp )=\int_{0}^{\beta
_{1}}p_{[\mathbf{L}]}^{z}(T_{e}^{(0)},\beta )\Phi _{z}(T_{e}^{(0)},\beta
)\,\,d\beta \, . \label{EQS}
\end{equation}
For each $z$, due to normalization conditions (\ref{normalization-1}), Eqs. (\ref{EQS}) may be considered as the Bayesian relations between the (partial) probability $P_{[\mathbf{L}]}^{z}$ for the photon emission in the peak $\mathbf{L}$ and the product of the conditional probability $p_{[\mathbf{L}]}^{z}(\beta)$ at the certain value of $\beta$ (local temperature) with the probability $\Phi _{z}(\beta )d\beta$ to find this value. Such interpretation of Equations (\ref{EQS}) allows to employ the relations of statistical physics, considering the values on the left side as the average distribution over random variable $\mathbf{L}$ and the $\Phi_{z}(\beta )$ as the most likelihood solution distribution over the random variable $\beta$.

This iterative procedure was applied to Eqs. (\ref{EQS}) providing the recurrent equality for $\Phi_z$-profiles:
\begin{equation}
\Phi _{z}^{N+1}(\beta )=\Phi _{z}^{N}(\beta
)\sum_{L}\frac{P_{[\mathbf{L}]}^{z}(\exp
)p_{[\mathbf{L}]}^{z}(T_{e}^{(0)},\beta )}{\int_{0}^{\beta
_{1}}\,p_{[\mathbf{L}]}^{z}(T_{e}^{(0)},\beta )\Phi _{z}^{N}\,(\beta
)d\beta }\, . \label{iteration}
\end{equation}
The values of the partial flux ratios for the peaks $(\Gamma_{\lbrack \mathbf{L}]}^{z})^{N}$ and that for the set of peaks $\mathbf{Z}$ chosen for the determination of the $\Phi
_{z}$-profile (``diagonal elements'') $({\Gamma _{\lbrack \mathbf{Z]}}^{z}})^{N}$ for the $N$-th iteration were calculated using Eq. (\ref{gamma}) and the set of equations followed from Eqs. (\ref{partial-gamma})
\begin{equation}
\Gamma _{[\mathbf{L}]}^{k}=\Gamma
_{[\mathbf{K}]}^{k}R_{\mathbf{Z}}^{\mathbf{K}}(\exp
)P_{[\mathbf{L}]}^{k},\,\, \mathrm{for}\,\, \mathbf{L}\subset
\mathbf{Z}\, ; \label{gammaN}
\end{equation}

\begin{equation}
\Gamma _{[\mathbf{He}]}^{he} = 1-\Gamma _{[\mathbf{H}]}^{h}R_{\mathbf{He}}^{\mathbf{H}}(\exp )P_{[\mathbf{He}]}^{h}\, ,  \label{gammaHe}
\end{equation}

\begin{equation}
\Gamma _{[\mathbf{\mathbf{Li}}]}^{li} = 1-
\Gamma_{[\mathbf{He}]}^{he}R_{\mathbf{Li}}^{\mathbf{He}}(\exp)P_{[\mathbf{Li}]}^{he}-\Gamma_{[\mathbf{H}]}^{h}R_{\mathbf{Li}}^{\mathbf{H}}(\exp )P_{[\mathbf{Li}]}^{h}\, ,  \label{gammaLi}
\end{equation}

\begin{eqnarray}
\Gamma _{[\mathbf{H}]}^{h} = \frac{1-R_{\mathbf{H}}^{\mathbf{He}}(\exp
)P_{[\mathbf{H}]}^{he}-R_{\mathbf{H}}^{\mathbf{Li}}(\exp
)P_{[\mathbf{H}]}^{li}+} {1-P_{[\mathbf{He}]}^{h}P_{[\mathbf{H}]}^{he}-
P_{[\mathbf{H}]}^{li}P_{[\mathbf{Li}]}^{h}+}                                \nonumber \\
\frac{+R_{\mathbf{H}}^{\mathbf{He}}(\exp
)P_{[\mathbf{H}]}^{li}P_{[\mathbf{Li}]}^{he}}{+P_{[\mathbf{He}]}^{h}P_{[\mathbf{Li}]}^{he}P_{[\mathbf{H}]}^{li}}\, ,  \label{gammaH}
\end{eqnarray}
where $\Gamma _{[\mathbf{Z}]}^{z}$ stands for $N$-th iteration and the measured values for the fluxes in the ranges $\mathbf{[Z]}$ were taken for the ratios $R_{\mathbf{Z}}^{\mathbf{K}}=F_{[\mathbf{K}]}^{\exp }/F_{[\mathbf{Z}]}^{\exp }$. The convergence was controlled by the $\chi ^{2}$ values:
\begin{equation}
\chi ^{2}=\sum_{[\mathbf{L}]}\left| \frac{F_{[\mathbf{L}]}^{\exp
}-F_{[\mathbf{L}]}}{F_{[\mathbf{L}]}^{\exp }}\right|^{2}.
\label{Kai}
\end{equation}
If radial profiles (\ref{radial profiles}) are known, it is more convenient to fulfill the iterative procedure for $\Phi_{z}(T_{e}^{(0)},\beta)$ in the ``$\rho$-representation''. In this case the equilibrium abundances $\widetilde{n}_{z}^{E}(\rho )$ may be
used as the zero approximation for $\Phi _{z}^{0}(T_{e}^{(0)},\beta )$.

The Bayesian inversion procedures have been iteratively applied to Eq. (\ref{EQS}) for $z$ from $li$ through $h$ ions with the following sets of peaks: {\bf He} = \{{\bf W},{\bf N}$_4$,{\bf N}$_3$,{\bf K}\}; {\bf Li} = \{{\bf Q,R,Z}\}; {\bf H} = \{{\bf X,Y,Z,W}\}. The parameter $T_{e}^{(0)}$ as well as the correction factors for atomic parameters from the set $\mathbf{D} $ have been found simultaneously minimizing $\chi ^{2}$ (Eq. (\ref{Kai})) and $\mid n(\beta )-1\mid$ (Eq. (\ref{sum})). It is worth to stress here that, since the condition (\ref{sum}) deals with the functions rather then with the single values (for example, the center values) bounding them in a wide range of $\beta $, it provides an efficient limitation on a possible solutions $\Phi_{z}^{Inv}(T_{e}^{(0)},\beta )$. Due to this fact some of the key parameters of the SM from the $\mathbf{D}$ set do not belong to the region of the optimized values $M_o$ and thus happens to be inconsistent with both Eqs. (\ref{gamma}), (\ref{flux}), and the condition (\ref{sum}) at any $P$, in particular the temperature $T_{e}^{(0)}$. In terms of the Bayesian inversion procedure used in this paper it means that there are no probable solutions for these parameters, providing the description of the measured spectra within the experimental error bars. Figures 5--7 show that these optimization conditions lead to rather strong limitation for $\Phi_z^{Inv}$-profile due to high sensitivity to 5\% variations of SM parameters, in particular, effective excitation rates $\mathbf{N}_3$ group of satellites, {\bf W} line.

\begin{figure}
\resizebox{0.45\textwidth}{!}{%
  \includegraphics[angle=270,scale=0.4]{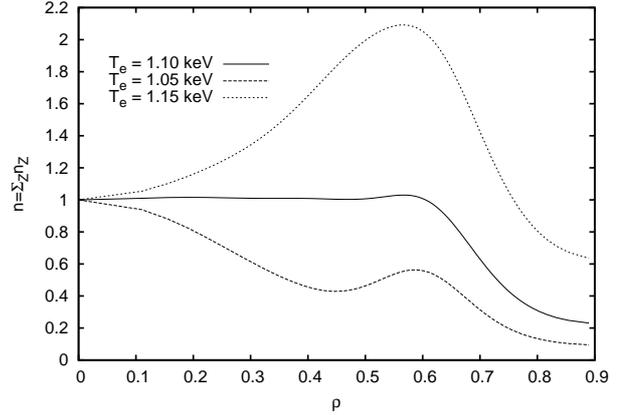}
}
\caption{Deviation of the sum of ionic abundances $n(\rho)$ from unity with 5\% variation of $T_{e0}$.}
\label{fig:5}
\end{figure}

\begin{figure}
\resizebox{0.45\textwidth}{!}{%
  \includegraphics[angle=270,scale=0.4]{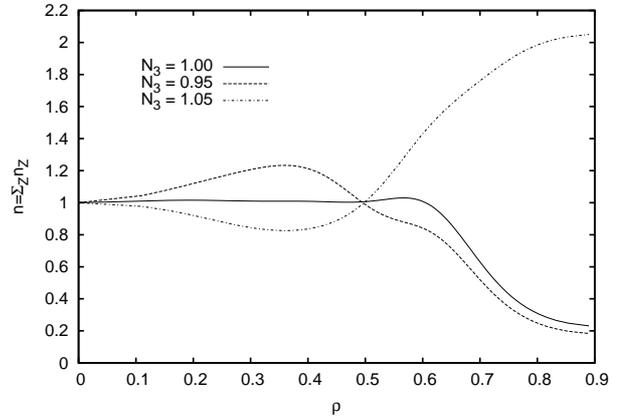}
}
\caption{Deviation of the sum of ionic abundances $n(\rho)$ from unity with variation of $\mathbf{N_3}$ dielectronic satellite group.}
\label{fig:6}
\end{figure}

\begin{figure}
\resizebox{0.45\textwidth}{!}{%
  \includegraphics[angle=270,scale=0.4]{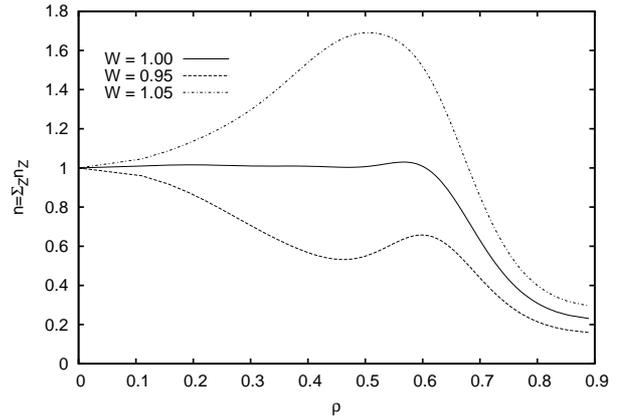}
}
\caption{Deviation of the sum of ionic abundances $n(\rho)$ from unity with variation of the excitation rate of $\mathbf{w}$ line.}
\label{fig:7}
\end{figure}

Thus studying the dependence of $\Phi_{z}^{Inv}$-profiles on their arguments, {\bf D} and $T_{e}^{(0)}$, it became possible to reveal by means of Eq. (\ref{sum}) the key parameters $\mathbf{D}$ and to establish the boundaries for their compatibility with the measured
spectra, i.e. the region $M_o$ \cite{Urn07}. The values of parameters which do not belong to the region $M_o$ are not self-consistent and hence are not correct. Our calculations showed that the solutions for $\Phi_z$-profiles are convex and are quite stable in the shape in respect to the random variations of key parameters of the SM. This result indicates that our inverse problem is not the ill-posed one and that the BIM is the regularization algorithm.

As was mentioned above a study of dependence of the functions $\Phi _{z}^{Inv}(\beta )$ on the arguments made it possible to determine the key parameters and the region $M_o$ of the consistency with the measured spectra. The set $\mathbf{D}=\{\alpha_l\}$ contains the ratios of effective excitation rates $\alpha_l = C_l^{he}/C_w^{he}$ for He-like ion lines {\bf x}, {\bf y}, {\bf z}, the satellite {\bf k} and the satellite group $\mathbf{N}_3$, and the set $\Phi = \{ T,\Phi_z(\beta)\}$; the region $M_o$ corresponds to the 5\% deviation of the quantities $\alpha_l(T)$ and $T$ from the optimized values. The factors $\gamma_l = \alpha_l / \alpha_l^{cor}$ are introduced to correct $\alpha_l(T)$ values.

\subsection{Iterative procedure}
\label{subsec:45}
Our study in the framework of the SCA includes a few iterative steps for the interpretation of the X-ray spectra under consideration using two complementary methods: the fitting procedure (FP) and the BIM inversion. The measured values for normalized radial profiles $\mathbf{T}_{e}(\rho )$ and $\mathbf{N}_{e}(\rho )$ are used in both cases. At the first step the FP is employed for determining the wavelengths, experimental fluxes in the peaks $F_{[\mathbf{L}]}^{\exp }$, and the core parameters $f_{i}$ (introduced for FP), utilizing the CIE values for radial profiles of ionic abundances $n_{z}^{(C)}(\rho )$. The values thus obtained are then used as the zero approximation for the next step where the radial profiles $n_{z}(\rho )$ are derived from the measured spectra by the BIM carried out simultaneously with the optimization procedure for the central temperature $T_{e}^{(0)}$ and the key atomic parameters $\mathbf{D}$. The values from BIM output are then applied as the incoming data for the FP and the calculation of synthetic spectra by the FP code. There are two criteria indicating the self-consistency of this iterative procedure: (i) the convergence of the results, and (ii) the coincidence of the corrected atomic data (within the experimental accuracy) for all the spectra measured at different conditions. The quantitative characteristics for the core approximation, $G_{\mathbf{L}_{2}}^{\mathbf{L}_{1}}$ factors defined above, are determined for various peaks at the final stage to check the sensitivity of the relative fluxes in the peaks to radial profiles.

\section{Verification and correction of atomic data}
\label{sec:5}

\subsection{Wavelength corrections}
\label{subsec:51}
In the first step of the spectra modeling, calculations of synthetic spectra were carried out according to the FP employing two sets of atomic data (LPI and PO). The wavelengths of the \textbf{w} and \textbf{z} lines were used as the reference points. Since the spectrometer on TEXTOR is set at an angle of about 10$^\circ$ relative to the toroidal direction of the tokamak, the wavelengths of lines were shifted on the detector due to the Doppler effect caused by plasma rotation. Thus, the position of the \textbf{w} line, $\lambda_{\mathbf{w}}$, was fitted to find the correct position of the spectra on the detector. The \textbf{w} and \textbf{z} lines have a different angle relative to the toroidal direction leading to slightly different shifts for each line; hence the distance between both lines depends on the plasma rotation and must be adjusted accordingly if the wavelengths are to remain in the correct relative position. To adjust for differential rotation, the dispersion was scaled linearly using a parameter $\mathbf{\alpha }$ determined by a fit of the line position of both lines.

In spite of the fact that the general shape of the synthetic spectra showed a good resemblance with the measured spectra, noticeable shifts in wavelengths of all prominent peaks (with the exception of the $\mathbf{X}$ and $\mathbf{Y}$ peaks and of the $\mathbf{N}_{3}$ and $\mathbf{N}_{4}$ satellite groups for the LPI data) was revealed for all measured spectra for both sets of atomic data, PO and LPI. In a next step new values for the wavelengths were obtained, therefore, by fitting calculated spectral profiles to each experimental peak varying the wavelengths of the corresponding lines in their vicinity as free parameters. These new wavelengths were then used in the synthetic spectra for the verification of atomic characteristics: $F_{2}$ factors for dielectronic satellites and effective rates $C^{eff}(T)$ for elementary processes as well as for accurate measurements of the flux intensity in the peaks $F_{[\mathbf{L}]}^{\exp }$ for the following BIM procedure. When the set of the most accurate atomic data needed for the determination of relative line intensities was selected after the next step, the procedure of wavelength determination was repeated till convergence. In fact, two iterations happened to be sufficient to achieve convergence with a required relative accuracy of $10^{-4}$. New values for the wavelengths (relative to the \textbf{w} line) thus obtained are given in Table 2 along with that measured by means of the EBIT devices \cite{Bei,Tar01}. The LPI values (see \cite{Gor06a}) were adopted for  the reference \textbf{w} and \textbf{z} lines. All absolute values of LPI and PO wavelengths for He-like ions were shifted to provide the same value for the \textbf{w} line given in \cite{Drake}
and \cite{Plante94}. A comparison of the TEXTOR and EBIT data shows a good agreement to within 0.3~m\AA\ for all wavelengths.

\begin{table*}
\begin{small}
\caption{Comparison of measured and calculated wavelength for Ar$^{16+}$ lines and their dielectronic satellites in the region 3.94--4.02~\AA .}
\begin{tabular}{ccccccccc}
\hline Key & Ref.\cite{Drake} & Ref.\cite{Plante94} & LPI & PO & Ref. \cite{ADAS} & Pres. work & Ref. \cite{Bei} & Ref.\cite{Tar01}  \\
\hline
\textbf{w} & 3.9491 & 3.9491 & 3.9491  & 3.9491 &        & 3.9491 & 3.9491 & 3.9491    \\
\textbf{x} & 3.9659 & 3.9658 & 3.9659  & 3.9661 &        & 3.9659 & 3.9658 & 3.9659    \\
\textbf{y} & 3.9694 & 3.9694 & 3.9693  & 3.9696 &        & 3.9693 & 3.9693 & 3.9694   \\
\textbf{z} & 3.9942 & 3.9941 & 3.9942  & 3.9949 &        & 3.9942 & 3.9942 & 3.9942   \\
\textbf{q} &        &        & 3.9813  & 3.9818 & 3.9754 & 3.9814 & 3.9813 & 3.9813  \\
\textbf{r} &        &        & 3.9834  & 3.9840 & 3.9779 & 3.9835 & 3.9834 & 3.9836   \\
\textbf{s} &        &        & 3.9676  & 3.9679 &        & 3.9676 & 3.9676 & 3.9689   \\
\textbf{t} &        &        & 3.9685  & 3.9688 &        & 3.9685 & 3.9685 & 3.9689   \\
\textbf{k} &        &        & 3.9898  & 3.9906 & 3.9849 & 3.9899 & 3.9899 & 3.9900   \\
\textbf{j} &        &        & 3.9939  & 3.9947 & 3.9891 &        & 3.9939 & 3.9939   \\
\textbf{a} &        &        & 3.9858  & 3.9663 &        &        & 3.9855 & 3.9857   \\
\textbf{m} &        &        & 3.9656  & 3.9660 &        &        & 3.9657 & 3.9658   \\
\hline
\end{tabular}
\end{small}
\end{table*}

\subsection{Analysis of relative flux intensities}
\label{subsec:52}

\subsubsection{Synthetic spectra and plasma parameters}
\label{subsubsec:521}
The results of calculations of the synthetic spectra fitted the experimental ones in the first step of the self-consistent modeling (with the CIE profiles for the relative ionic abundances $\widetilde{n}_{z}^{C}(\rho )$) for seven arbitrary selected discharges with different core temperatures $T^{\ast}$ (measured by the ECE method) and densities are presented in Table 3. Three rows for each spectrum in this table corresponds, respectively, to the LPI, PO, and ADAS atomic data banks used for calculations. The core plasma parameters $f=\{f_{e}$, $f_{li}$, $f_{h}\}$ optimized by the FP as well as the $E$-ratios (experimental-to-predicted values of the peak fluxes), and $G_{\mathbf{\sigma }_{2}}^{\mathbf{\sigma }_{1}}$-factors calculated due to Eq. (\ref{G-L1-L2-factor}) are given for the relative flux intensities of the peaks $\mathbf{L}$ identified above in the section of spectra description.

\begin{table*}
\begin{small}
\caption{Comparison of observed-to-calculated intensities.}
\scriptsize
\begin{tabular}{cccccc|cccccc|ccccc}
\hline
\multicolumn{6}{c|}{Plasma parameters }&\multicolumn{6}{|c|}{ E = Experiment/Theory }&\multicolumn{5}{|c}{ G-ratio } \\ \hline
\# & $N_e$,  & $T^{\ast}$, & $T_{e0}$, & $\frac{\widetilde{n}_{H}}{\widetilde{n}_{H}^C}$ &
$\frac{\widetilde{n}_{Li}}{\widetilde{n}_{Li}^C}$ & $N_3$/W & X/W & Y/W & K/W & Z/W & $R_2$ &
$N_3$/W & K/W & X/W & Z/W & $G_2$ \\
  & cm$^{-3}$ & keV & keV & & & & & & & & & & & & & \\
\hline 1 & 4.8$\cdot$10$^{13}$ & & & & & & & & & & & & & & & \\
\hline
  &             &    & 1.15  & 1.53 & 1.89 & 0.87 & 0.99 & 1.02 & 1.02 & 0.98 & 1.02 & 1.25 & 1.4 & 1.06 & 1.06 & 1.05  \\
2 & 5.4$\cdot$10$^{13}$ & 1.13  &  1.14 & 1.67 & 1.77 & 0.88 & 0.95 & 1.0 & 1.03 & 0.99 & 0.99 & 1.24 & 1.36 & 1.05 & 1.02 & 1.02 \\
  &  & & 1.15 & 1.60 & 2.49 & 0.92 & 0.97 & 1.03 & 1.04 & 1.04 & 0.98 & 1.25 & 1.37 & 1.11 & 1.10 & 1.05 \\ \hline
  &  &  & 1.24 & 1.35 & 2.0 & 0.99 & 1.08 & 1.04 & 1.0 & 0.97 & 1.04 & 1.2 & 1.23 & 1.05 & 1.05 & 1.06 \\
3 & - & 1.15 & 1.26 & 1.23 & 1.98 & 0.97 & 1.02 & 1.02 & 0.99 & 0.99 & 1.03 & 1.2 & 1.17 & 1.06 & 1.06 & 1.05 \\
 & & & 1.22 & 0.72 & 2.35 & 0.97 & 1.11 & 1.01 & 1.00 & 1.01 & 1.04 & 1.19 & 1.24 & 1.08 & 1.07 & 1.07 \\ \hline
 & & & 1.25 & 0.63 & 1.90 & 0.94 & 1.05 & 1.09 & 1.03 & 0.99 & 1.08 & 1.36 & 1.55 & 1.11 & 1.12 & 1.18 \\
4 & 5.6$\cdot$10$^{13}$ & 1.24 & 1.21 & 0.49 & 1.73 & 0.94 & 1.02 & 1.08 & 1.0 & 1.0 & 1.06 & 1.36 & 1.54 & 1.12 & 1.11 & 1.19 \\
 & & & 1.22 & 0.24 & 2.40 & 0.93 & 1.0 & 1.03 & 1.0 & 0.97 & 1.03 & 1.34 & 1.54 & 1.16 & 1.12 & 1.15 \\ \hline
 & & & 1.22 & 0.58 & 1.99 & 0.94 & 1.03 & 1.06 & 1.02 & 0.99 & 1.05 & 1.28 & 1.44 & 1.09 & 1.10 & 1.09 \\
5 & 4.1$\cdot$10$^{13}$ & 1.43 & 1.20 & 0.96 & 1.86 & 0.94 & 0.99 & 1.08 & 1.04 & 0.98 & 1.04 & 1.27 & 1.43 & 1.09 & 1.08 & 1.08 \\
 & & & 1.23 & 0.17 & 2.59 & 0.96 & 1.00 & 1.04 & 1.03 & 1.01 & 1.01 & 1.27 & 1.43 & 1.12 & 1.12 & 1.12\\ \hline
 & & & 1.62 & 0.91 & 2.37 & 0.91 & 0.96 & 1.06 & 1.03 & 1.0 & 1.02 & 1.29 & 1.4 & 1.08 & 1.07 & 1.09 \\
6 & - & 1.65 & 1.66 & 0.96 & 2.44 & 0.93 & 0.94 & 1.07 & 1.06 & 1.0 & 1.02 & 1.3 & 1.41 & 1.09 & 1.07 & 1.08 \\
 & & & 1.64 & 0.53 & 3.31 & 0.98 & 1.00 & 1.06 & 1.03 & 0.99 & 1.05 & 1.31 & 1.51 & 1.12 & 1.10 & 1.11 \\ \hline
 & & & 2.77 & 0.97 & 5.08 & 1.06 & 0.99 & 1.07 & 1.03 & 1.01 & 1.03 & 1.74 & 2.13 & 1.20 & 1.16 & 1.17 \\
7 & 1.2$\cdot$10$^{13}$ & 2.14 & 2.7 & 0.8 & 4.89 & 0.99 & 0.99 & 1.11 & 1.0 & 1.0 & 1.04 & 1.65 & 1.69 & 1.17 & 1.12 & 1.13 \\
 & & & 2.63 & 0.76 & 6.66 & 0.98 & 0.98 & 1.08 & 1.0 & 1.0 & 1.02 & 1.66 & 1.70 & 1.19 & 1.14 & 1.15 \\ \hline
\end{tabular}
\end{small}
\end{table*}

\begin{table*}
\begin{small}
\caption{Comparison of observed-to-calculated intensities.}
\scriptsize
\begin{tabular}{cccccc|cccccc|ccccc}
\hline
\multicolumn{6}{c|}{Plasma parameters }&\multicolumn{6}{|c|}{ E = Experiment/Theory }&\multicolumn{5}{|c}{ G-ratio } \\ \hline
\# & $N_e$, & $T^{\ast}$, & $T_{e0}$, & $\frac{\widetilde{n}_{H}}{\widetilde{n}_{H}^C}$ &
$\frac{\widetilde{n}_{Li}}{\widetilde{n}_{Li}^C}$ & $N_3$/W & X/W & Y/W & K/W & Z/W & $R_2$ &
$N_3$/W & K/W & X/W & Z/W & $G_2$ \\
  & cm$^{-3}$ & keV & keV & & & & & & & & & & & & & \\
\hline
 & & & & & & & & & & & & & & & & \\
1 & 4.8$\cdot$10$^{13}$ & 0.9 & 0.975 & 0.78 & 2.07 & 0.96 & 1.03 & 1.01 & 1.0 & 0.99 &  1.03 & 1.21 & 1.35 & 1.07 & 1.09 & 1.10 \\
 & & & 0.975 & 0.50 & 1.81 & 0.96 & 1.04 & 1.01 & 1.0 & 0.99 & 1.03 & 1.22 & 1.35 & 1.06 & 1.10 & 1.09 \\ \hline

& & & 1.16 & 0.8 & 1.83 & 0.93 & 1.01 & 1.03 & 1.0 & 1.0 & 1.01 & 1.25 & 1.40 & 1.07 & 1.08 & 1.08 \\
2 & 5.4$\cdot$10$^{13}$ & 1.13  & 1.1 & 0.8 & 1.67 & 0.95 & 1.02 & 1.03 & 1.0 & 0.98 & 1.04 & 1.2 & 1.33 & 1.12 & 1.16 & 1.15 \\
  &  & &  1.11 & 0.60 & 1.85 & 0.95 & 1.03 & 1.01 & 1.0 & 0.99 & 1.03 & 1.21 &  1.34 & 1.09 & 1.13 & 1.12 \\ \hline

  &  & & 1.19 & 0.57 & 1.83 & 0.96 & 1.03 & 1.08 & 1.0 & 1.0 & 1.06 & 1.19 & 1.31 & 1.06 & 1.07 & 1.07 \\
3 & - & 1.15 & 1.17 & 0.89 & 1.96 & 0.96 & 0.96 & 1.08 & 1.0 & 0.98 & 1.05 & 1.15 & 1.24 & 1.12 & 1.15 & 1.14\\
 & & & 1.17 & 0.39 & 1.62 & 0.95 & 1.04 & 1.02 & 1.0 & 0.98 & 1.05 & 1.15 & 1.24 & 1.11 & 1.13 & 1.12\\ \hline
 & & & 1.24 & 0.45 & 1.86 & 1.05 & 1.06 & 1.10 & 1.0 & 1.0 & 1.08 & 1.36 & 1.55 & 1.11 & 1.12 & 1.12 \\
4 & 5.6$\cdot$10$^{13}$ & 1.24 & 1.19 & 0.39 & 1.41 & 1.05 & 1.0 & 1.06 & 1.0 & 0.98 & 1.06 & 1.29 & 1.45 & 1.23 & 1.28 & 1.26\\
 & & & 1.20 & 0.45 & 1.69 & 1.04 & 1.05 & 1.03 & 1.0 & 0.97 & 1.07 & 1.33 & 1.49 & 1.13 & 1.17 & 1.15\\ \hline
 & & & 1.22 & 0.6 & 1.95 & 1.03 & 1.04 & 1.07 & 1.0 & 1.0 & 1.06 & 1.28 & 1.43 & 1.09 & 1.10 & 1.09\\
5 & 4.1$\cdot$10$^{13}$ & 1.43 & 1.15 & 0.75 & 1.75 & 1.01 & 1.0 & 1.03 & 1.0 & 0.99 & 1.04 & 1.24 & 1.38 & 1.13 & 1.17 & 1.15\\
 & & & 1.15 & 0.33 & 1.76 & 1.01 & 1.04 & 1.01 & 1.0 & 0.99 & 1.04 & 1.25 & 1.38 & 1.09 & 1.13 & 1.12\\ \hline
 & & & 1.62 & 0.8 & 2.5 & 0.99 & 0.97 & 1.09 & 1.0 & 1.0 & 1.04 & 1.29 & 1.42 & 1.08 & 1.07 & 1.07\\
6 & - & 1.65 & 1.54 & 1.19 & 2.58 & 1.02 & 0.89 & 1.08 & 1.0 & 1.0 & 1.0 & 1.28 & 1.43 & 1.06 & 1.07 & 1.07\\
 & & & 1.55 & 0.36 & 2.44 & 1.02 & 0.97 & 1.01 & 1.0 & 1.0 & 0.99 & 1.28 & 1.42 & 1.06 & 1.07 & 1.07 \\ \hline
 & & & 2.55 & 0.70 & 4.67 & 1.05 & 1.00 & 1.09 & 1.0 & 1.0 & 1.04 & 1.59 & 1.92 & 1.17 & 1.14 & 1.15\\
7 & 1.2$\cdot$10$^{13}$ & 2.14 & 2.13 & 0.96 & 4.22 & 1.03 & 0.91 & 1.08 & 1.0 & 0.99 & 1.02 & 1.42 & 1.58 & 1.08 & 1.11 & 1.08\\
 & & & 2.13 & 0.14 & 4.06 & 1.03 & 0.98 & 1.02 & 1.0 & 1.0 & 1.0 & 1.42 & 1.57 & 1.05 & 1.06 & 1.06\\ \hline
\end{tabular}
\end{small}
\end{table*}

The ratios introduced by Gabriel \& Jordan \cite{Gab69}, denoted here as $R_{2}=(\mathbf{X}+\mathbf{Y})/\mathbf{Z}$ and $G_{2}=(\mathbf{X+Y+Z})/ \mathbf{W}$, are also presented. For the spectrum 1 the FP failed to optimize the model parameters with the equilibrium ionic abundances, because of the negative values for the parameter $f_{h}$ needed to fit the experimental ratio $\mathbf{Z}/\mathbf{W}$.

It is worth noting here that the FP, based on the optimization of the spectral intensity in the spectral regions [$\mathbf{L}$], does not provide the optimization (equality) of the corresponding fluxes in these regions and hence the $E$-ratios for $\mathbf{K}/\mathbf{W}$ and $\mathbf{Z}/\mathbf{W}$ values are not equal to unity. The analysis showed that this effect, leading to noticeable errors in the temperature and H-like ion abundances determination, is mainly connected with the distortion of the line shape at the ``wings'' of line spectral profiles caused by the apparatus function and with the uncertainty in the background determination. The corrected values for plasma parameters, $E$- and $G$-ratios obtained by more accurate account for the contribution of line profiles (``wings'' of neighboring lines contributing to the factors $\pounds_{[\mathbf{L}]}^{l}(T_{i})$ defined by Eq. (\ref{shape-correction}) to the peak fluxes, are given in Table 4 for the LPI data (first rows for each spectrum). This correction was made by comparing the measured and the calculated fluxes in the spectral bands where the ``wings'' of the most prominent lines were not affected by other lines (for example short wavelength region of the \textbf{w} line and the long wavelength region of the \textbf{z} line). Since, as is seen from these results, the effect of corrections is quite noticeable for the $f_{h}$ and the $f_{li}$ values (in particular at low temperatures), it is worth to compare only the relative results obtained within the same approach for different sets of atomic data (see Table 3). This comparison shows that all three sets of $AD$ provided the agreement of the relative peak intensities within the accuracy of the measurements $\simeq$10\%, besides the $\mathbf{R}/\mathbf{W}$ and $\mathbf{K}/\mathbf{N}_{3}$ ratios; disagreement for the latter ratio, being the function of the core temperature, exceeded this value for two spectra 2 and 6. The deviation of the $\mathbf{R}/\mathbf{W}$ ratio as was shown in \cite{Mar04} is connected with cascades from autoionization states which were ignored in the previous and the present calculations; the account for this effect in the BIM procedure leads to the E-ratio close to unity as is seen from Table 4.

The values of the core temperatures are very close for all the sets of $AD$; the LPI and PO data provides close values for $f_{z}$ parameters while the ADAS set gives larger values for $f_{li}$ and smaller values for $f_{h}$. It is also worth to note that although the $E$-ratios for some values of relative fluxes ($\mathbf{N}_{3}/\mathbf{W}$ , $\mathbf{Y/W}$, $R_{2}$) are within the error bars for all the data sets, the deviation of the calculated values from the measured ones of about 4--10\% has the same tendency for all spectra with the temperature larger 1.15~keV.

The calculations made in the second step of the modeling by means of the BIM showed the deviation of the profiles of the ionic abundances from the equilibrium. The non-equilibrium profiles thus obtained were used for the calculations of the synthetic spectra by the FP code accounting for the aforementioned corrections of the line profiles as well as the corrections of the atomic data (LPI) followed from the optimization made along with the BIM. The results of the BIM calculations are given in Table 4 for the following corrections: for each discharge theoretical values of effective excitation rates were multiplied on the correction factors for {\bf z} line (the second row) and for {\bf y} line (the third row) in order to improve the observed-to-predicted ratio $E$ for the $R_{2}$ values. The corrected values for $F_{2}$ factors for the intensities of dielectronic satellites were also used (all rows); the resulted values for these factors are given in Table 1 (column ``Pres. work''). These values were obtained by fitting one variable parameter for all satellites -- effective charge $Z_{eff}$, used for calculations of corresponding autoionization widths. In the paper \cite{Gor06a} this charge equals to the spectroscopic symbol of Li-like argon ion (LPI values in Table 1). In the present work the best fit of all experimental spectra was achieved by $F_2$ factors with $Z_{eff}$=15. The comparison of peak intensities is presented below in the second section; here we will pay attention to the plasma parameters only.

As is seen from Table 4, there is a considerable decrease of the $f_{li}$ values obtained from the BIM results (in particular with corrected $E$-ratio for the {\bf y} line) in comparison to the FP; the core temperatures $T_{e0}$ became in better agreement (apart of the spectrum 5) with $T^{\ast}$ measured by the ECE method, especially for the hottest spectrum 7. These changes are associated with the deviation of radial abundance profiles from equilibrium as is indicated by the $G_{\mathbf{W}}^{\mathbf{K}}$ factors. The changes of plasma parameters resulting from the corrections of $AD$ indicate also their self-consistent character. For the spectrum 5 both Tables 3 and 4 present a difference of about 17\% showing that the reason of this difference is not related to the distortion of ionic abundances or a numerical error; hence we are left with no other option than to ascribe this error to the value of $T^{\ast }$. The important results concerns the relative ionic abundances at the plasma core. The values of the $f_{h}$ parameter in both Tables 3 and 4 are $<1$; this can be explained by two effects: ion transport and charge transfer from neutral atoms of the working gas -- neutral hydrogen atoms \cite{Mar06c}. Good agreement (accounting for an error in the determination of $f_{h}$ and $f_{li}$ parameters which we estimated as being of $\simeq $ 30\% and 20\% , respectively) with the quantitative prediction of the latter effect made in \cite{Mar06b} with the help of impurity transport model (ITM) was obtained for all discharges with the central values of the density of neutrals varying in the range (1--5)$\cdot$10$^{7}$~cm$^{-3}$; these values and the radial profiles for these densities were adopted from the calculations by the ITM.

The radial profiles of ionic abundances $n_{z}(\rho )$ have been compared with those calculated in the framework of the ITM including the effects of transport of argon ions and charge transfer \cite{Mar06b,Mar06c}. These calculations based on Eqs. (\ref{basic equations}) with the semi-empirical diffusion coefficients revealed good qualitative agreement with the BIM results. The comparison of the central values for $\widetilde{n}_{z}(T_{e0})$ are shown in Fig. 8. The detailed quantitative analysis of the plasma parameters obtained within the SM is out of the scope of the present paper and will be given elsewhere. Here we have only to remark that these parameters happened to be rather sensitive to the core temperature $T_{e0}$. Figure 7 demonstrates the deviation of the sum of ionic abundances $n(\rho )$ from unity with 5\% variations of $T_{e0}$. Thus the determination of $T_{e0}$ by the optimization of the equality (\ref{sum}) provides a new method of its accurate diagnostics.

\begin{figure}
\resizebox{0.5\textwidth}{!}{%
  \includegraphics{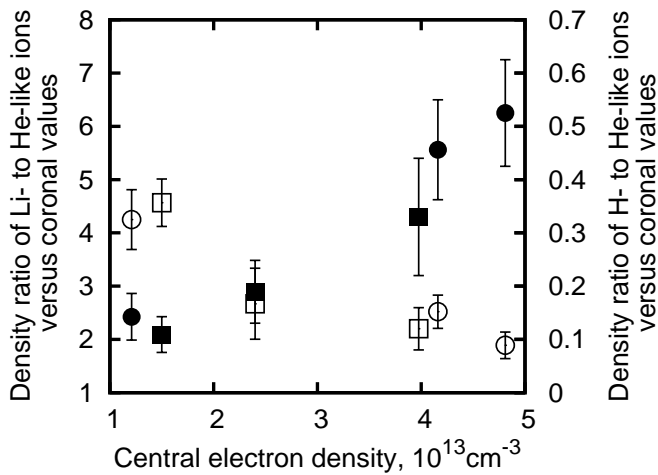}
}
\caption{Comparison of the central values for the relative values (to $z=He$) ionic
abundances $\widetilde{n}_{z}(T_{e0})$ obtained in the frame of ITM and SM. Density of the Li-like ions, obtained using ITM is demonstrated with open squares and the density of H-like ions with filled squares. Density of the Li-like ions, obtained using SM is demonstrated with open circles and the density of H-like ions with filled circles.}
\label{fig:8}
\end{figure}

\subsubsection{Correction of atomic data for the He-like ion}
\label{subsubsec:522}
The $\mathbf{W,X,Y}$, and $\mathbf{Z}$ peaks formed mainly by corresponding He-like ion lines $\mathbf{w,x,y,}$, and $\mathbf{z}$ have the most complex spectral structure since they are blended with a series of dielectronic satellite lines emitted by Li-like ions, which converge on them. Furthermore, the flux intensity of He-like ion lines includes contributions from recombination of H-like ions, from ionization of Li-like ions ({\bf z} line) and from cascades from higher levels. The remarkable feature is that the calculated (with all data sets) $R_{2}$ ratios practically do not depend on the electron temperature in the range 1--2 keV while the flux ratio of corresponding lines $i_{\mathbf{z}}^{\mathbf{x+y}}(T)$ reveal a weak dependence within about 5\%. This specific for argon ions ``accidental'' compensation of the $T_e$-dependence in $R_{2}=i_{\mathbf{z}}^{\mathbf{x}+\mathbf{y}}(T_e)\phi (P)$ values by the aforementioned processes, characterized by the correction factor $\phi (P)$, results in the equality to unity (within 3\%) of the $G$-factor for the $R_{2}$ ratio (see Tables 3 and 4). In that case the $(\mathbf{X+Y)}/\mathbf{Z}$ flux ratio depends only on the core values of plasma parameters $P$; on the other hand, since the factor $\phi (T_e)$, being close to unity, is practically independent on the ionic abundances, this ratio happens to be independent on all parameters of the model. Thus the comparison of the $R_{2}$-ratio with the experimental values provides a unique possibility to verify the accuracy for the flux ratio of lines $i_{\mathbf{z}}^{\mathbf{x+y}}(T)$ proportional to the ratio of corresponding rates. Both Tables 3 and 4 show that the deviation of the predicted values from the measured ones is within 6\% for all the sets of atomic data and all the spectra, besides the LPI, giving 8\% for the spectrum 4. At the same time the predicted $\mathbf{Y}/\mathbf{W}$ flux ratio for all the data set has the tendency to underestimate (up to $\simeq $ 10\%) and the $\mathbf{Z}/\mathbf{W}$ ratio to overestimate (up to $\simeq $3\%) the experimental one. The correction made by the optimization of effective rate ratios for $\mathbf{y}$ and $\mathbf{w}$ lines shows that the $\mathbf{Y}/\mathbf{W}$ flux became very close to the experimental one (see the third row in Table 4).

\subsubsection{Correction of atomic data for Li-like ion}
\label{subsubsec:523}
In distinction to the He-like ion lines the intensities of the dielectronic satellites emitted by the Li-like ions directly proportional to $F_{2}$ factors are measured and predicted with less accuracy. As follows from Table 1, where the results of calculations for $F_{2}$ by various methods are given, the deviation is of about 10\% for intensive satellites and up to 30\% for low intensities. The most noticeable difference is seen between the LPI and the PO calculations for autoionization probabilities $A_{a}$. Note that the overestimation of the $\mathbf{N}_{3}$ and $\mathbf{N}_{4}$ satellites obtained by the MZ code was also a subject of discussion in connection with the spectra measured at the NSTX \cite{Bit03}. This disagreement was shown to be connected with the effect of screening of the optical electron by other electrons which was not accounted for previously for $A_{a}$ values calculated with hydrogen-like functions. In the paper \cite{Gor06a} this effect was taken into account by introducing the basis functions calculated by the ATOM program to the MZ code. This procedure substantially decreased the autoionization widths and thus corresponding $F_{2}$ factors (denoted as LPI) which became close to the PO and ADAS values (see Table 1).

The comparison of ratios of the $F_{2}$ factors for the $\mathbf{k}$ and $\mathbf{N}_{3}$ satellites provided by different methods with the optimized value derived from the TEXTOR spectra showed that all sets of calculated data overestimate the latter by about 8, 10, and 13\%, respectively, for the ADAS, LPI, and PO sets.

\section{Discussion and conclusions}
\label{sec:6}
Summarizing the results of analysis of the observed-to-predicted ratios for line intensities one may conclude that all three sets of atomic data used in the present paper provide the same accuracy for the synthetic spectra for the temperature range 0.9--2.5~keV. The ADAS data gives better values for the $\mathbf{N}_{3}$ group of dielectronic satellites emitted by Li-like ions, however all the data needs the correction for 5--10\%. All the sets revealed the deviation of the relative effective rates for $\mathbf{y}$ line from the experimental values up to about 10\% at the temperatures of the core $T_{e0}>\sim$~1.2~keV, while other lines emitted by the He-like ions are in a good agreement within $\simeq$~5\%.

The application of the BIM, which revealed the deviation of the radial profiles for the ionic abundances, showed that the account of this effect is important at large temperatures of the core $T_{e0}>\sim$~1.5~keV; for the temperatures less or $\simeq$~1~keV the account of the Be-like ions became necessary. Good agreement of ionic abundances obtained by the BIM in the framework of the spectroscopic model (SM) and the
impurity transport model (ITM) based on the numerical modeling justifies the accuracy of the approach. It is also worth to note that the analysis of the accuracy of the LPI data used for a description of $K_{\beta }$ spectra in \cite{Gor03} supports the conclusions of the present paper and the self-consistency of the spectroscopic model.

Finally we come to the following general conclusions. The possibility have been firstly shown to use the tokamak TEXTOR, due to its features and unique equipment with diagnostics techniques, for precision measurements and verification of appropriate atomic data by means of the X-ray spectroscopy. The spectra obtained by means of the Bragg spectrometer/polarimeter with high spectral, spatial and temporal resolution have been used to measure the wavelengths of main Ar$^{16+}$ and Ar$^{15+}$ lines with a relative accuracy of 10$^{-4}$. A new approach based on self-consistent modeling in the frame of the SM was justified and applied to atomic data analysis by means of a set of $K$-spectra of argon ions in the range 3.94--4.02 \AA\ firstly recorded for a wide range of the central temperatures with high accuracy within 10\%. Two complimentary procedures, fitting procedure and Bayesian inversion, were used iteratively to derive all identified model parameters and to provide the synthetic spectra resembling the experimental ones within the experimental errors.

The measurements of wavelengths with a relative accuracy of 10$^{-4}$ and the main collisional characteristics with an accuracy of 5--10\% fulfilled by means of this approach became possible due to a set of spectra firstly recorded with high accuracy within 10\% for a wide range of electron temperatures of the tokamak core 0.9--2.5~keV. The TEXTOR tokamak equipment containing a unique set of various diagnostic instrumentation helped to justify the self-consistency of modeling and the accuracy for spectroscopically measured plasma parameters. The approach provided also a new method for the determination of the plasma core temperature with high accuracy to within 5\%.

The modified atomic data calculated by two method based on the multi-configuration and the perturbation theory expansions have been used in the final step of modeling in order to check their accuracy and the self-consistency of the model. A wide range of conditions covered in the TEXTOR tokamak made it possible to apply the ``stimulated selection'' method providing the verification of characteristics of elementary processes in coronal plasma with a relative accuracy of the measurements of 5--10\%. The results of diagnostics obtained for the central electron temperatures are in a good agreement to within about 5\% with the values measured by the ECE method confirming the accuracy of both measurements.

Such a high precision in the determination of atomic characteristics, 4--6 times exceeding the results of previous experimental and theoretical studies, besides the fundamental importance for the physics of highly ionized ions, is a necessary condition for application of advanced methods of spectroscopic diagnostics to investigation of hot laboratory and astrophysical plasmas, in particular non-steady-state phenomena. Further two-dimensional studies of He-like ion spectra \cite{Ber04b} have to expand the ranges for the verification and diagnostics abilities of the X-ray spectroscopy method.

\begin{acknowledgement}
The author is grateful to Dr. M.~Bitter for establishing X-ray spectroscopy on TEXTOR tokamak, Dr. O.~Marchuk and Dr. G.~Bertschinger for providing experimental data
and helping in their processing, Prof. H.-J.~Kunze for supporting and fruitful discussions, Prof. L.~Vainshtein and Dr. J.~Dubau for helping with atomic data calculations, Dr. P.~Beiersdorfer for providing with EBIT data, Prof. R.~Janev for consultations and the help in calculations of charge transfer cross-sections, and S.~Oparin for an important assistance in the application of the BIM.
\end{acknowledgement}

\begin{appendix}

\def\thesection{} 

\section{Appendix. Bayesian iterative scheme}

\def\thesection{A}

In order to formulate the Bayesian iterative method, consider two related complete systems of events $\{X_i\}$ and $\{Y_k\}$ ($i=1,\dots,n$; $k=1,\dots,m$) as well as corresponding probability distributions $\{P(X_i)\}$ and $\{P(Y_k)\}$ for them. In applications these distributions may also be ones for some random variables $\mathbf{X}$ and $\mathbf{Y}$. Let the probability distributions $\{P(X_i)\}$ and $\{P(Y_k)\}$ be related by the formulas of the total probability:
$$
P(Y_k) = \sum_i P(Y_k|X_i)\, P(X_i) \, , \eqno (\mathrm{A}.1)
$$
where $P(Y_k|X_i)$ is the conditional probability of the event $Y_k$ at the condition $X_i$. If the distribution $\{P(Y_k)\}$ is known, one can formulate the problem for deriving the $\{P(X_i)\}$ one from the relations (A.1). Below we will state an iterative procedure called the BIM to resolve this task.

The BIM is based on Bayes' theorem for the a posteriori conditional probability connecting two random variables defined on the fields of events $\{X_i\}$ and $\{Y_k\}$ as follows:
$$
P(X_i|Y_k) = \frac{P(Y_k|X_i)\, P(X_i)}{\sum\limits_j P(Y_k|X_j)\, P(X_j)} \, . \eqno (\mathrm{A}.2)
$$
The formula of the total probability for the distribution $P(X_i)$ is then given by the expression inverse to (A.1):
$$
P(X_i) = \sum_k P(X_i|Y_k)\, P(Y_k) \, . \eqno (\mathrm{A}.3)
$$
Substituting (A.2) in (A.3) gives the identity
$$
P(X_i) = P(X_i)\, \sum_k \frac{P(Y_k|X_i)\, P(Y_k)}{\sum\limits_j P(Y_k|X_j)\, P(X_j)} \, . \eqno (\mathrm{A}.4)
$$
The expression (A.4) can be used for formulating an iterative scheme. For this purpose, the value $P(X_i)$ in the right side of (A.4) is interpreted as step $n$, and in the left side as step $(n+1)$ of the iterative procedure. Thus, one obtains the following recurrence relation for the distribution $P(X_i)$:
$$
P^{(n+1)}(X_i) = P^{(n)}(X_i)\, \sum_k \frac{P(Y_k|X_i)\, P(Y_k)}{\sum\limits_j P(Y_k|X_j)\, P^{(n)}(X_j)} \, . \eqno (\mathrm{A}.5)
$$
It is worth to make some comments regarding the formula (A.5). The left side of (A.5) can be considered as the estimate of the $n$-th hypothesis for the probability distribution $P(X_i)$. The initial approximation for a priori distribution $P^{(0)}(X_i)$ may be taken in accordance with any prior information. If such information is absent, according to the Bayes' postulate, one assumes a uniform distribution (corresponding to equal lack of knowledge). It is also possible, using relation (A.5), to show that the normalizing condition for the distribution $\{P(X_i)\}$ is automatically conserved at any step of the iterative procedure (A.5).

\end{appendix}

%
%

\end{document}